\documentclass[preprint,article,accept,moreauthors,pdftex]{Definitions/mdpi} 

\firstpage{1} 
\pubvolume{1}
\issuenum{1}
\articlenumber{0}
\pubyear{2021}
\copyrightyear{2020}
\datereceived{} 
\dateaccepted{} 
\datepublished{} 
\hreflink{https://doi.org/} 

\Title{Promise of persistent multi-messenger astronomy with the blazar OJ~287}

\TitleCitation{Promise of persistent multi-messenger astronomy with the blazar OJ~287}

\Author{Mauri J. Valtonen $^{1,2}$\orcidA{}, Lankeswar Dey $^{3}$\orcidB{}, A. Gopakumar$^{3}$\orcidC{}, Staszek Zola$^{4}$\orcidD{}, S. Komossa$^{5}$\orcidE{}, 
Tapio Pursimo$^{6}$, Jose L. Gomez$^{7}$\orcidF{}, Rene Hudec$^{8,9}$\orcidG{}, Helen Jermak$^{10}$ and Andrei V. Berdyugin$^{2}$}

\AuthorNames{Mauri Valtonen, Lankeswar Dey, A. Gopakumar, Staszek Zola, S. Komossa, Tapio Pursimo, Jose L. Gomez, Rene Hudec, Helen Jermak and Andrei Berdyugin}

\AuthorCitation{Valtonen, M.; Dey, L.; Gopakumar, A.; et al.}

\address{%
$^{1}$ \quad FINCA, University of Turku, Turku, Finland; mvaltonen2001@yahoo.com\\
$^{2}$ \quad Tuorla Observatory, Department of Physics and Astronomy, University of Turku, Turku, Finland\\
$^{3}$ \quad Department of Astronomy and Astrophysics, Tata Institute of Fundamental Research, Mumbai, India\\
$^{4}$ \quad Astronomical Observatory, Jagiellonian University, ul. Orla 171, 30-244 Krakow, Poland\\
$^{5}$ \quad Max-Planck-Institut f\"ur Radioastronomie, Auf dem H\"ugel 69, 53121 Bonn, Germany\\
$^{6}$ \quad Nordic Optical Telescope, Apartado 474, E-38700 Santa Cruz de La Palma, Spain\\
$^{7}$ \quad Instituto de Astrofisica de Andalucia - CSIC, Glorieta de la Astronomia s/n, 18008 Granada, Spain\\
$^{8}$ \quad Czech Technical University, Faculty of Electrical Engineering, Prague, Czech Republic\\
$^{9}$ \quad Kazan Federal University, Kazan, Russian Federation\\
$^{10}$ \quad Astrophysics Research Institute, Liverpool John Moores University, Liverpool Science Park IC2, 146 Brownlow Hill, UK\\ }

\corres{Correspondence: mvaltonen2001@yahoo.com, lanky441@gmail.com}

\abstract{Successful observations of the seven predicted bremsstrahlung flares from the unique bright blazar OJ~287 firmly point to the presence of a nanohertz gravitational wave (GW) emitting supermassive black hole (SMBH) binary central engine. We present arguments for the continued monitoring of the source in several electromagnetic windows to firmly establish various details of the SMBH binary central engine description for OJ~287. In this article, we explore what more can be known about this system, particularly with regard to accretion and outflows from its two accretion disks. We mainly concentrate on the expected impact of the secondary black hole on the disk of the primary on December 3, 2021 and the resulting electromagnetic signals in the following years. We also predict the times of exceptional fades, and outline their usefulness in the study of the host galaxy. A spectral survey has been carried out, and spectral lines from the secondary were searched for but not found. The jet of the secondary has been studied and proposals to discover it in future VLBI observations are mentioned. In conclusion, the binary black hole model explains a large number of observations of different kinds in OJ~287. Carefully timed future observations will be able to provide further details of its central engine. Such multi-wavelength and multidisciplinary efforts will be required to pursue multi-messenger nanohertz GW astronomy with OJ~287 in the coming decades.}


\begin{document}

\section{Introduction}

Supermassive black hole (SMBH) binary systems are expected to be common in the Universe \cite{BBR80,val89,mik92,val96b,qui96,mil01,vol03}.
Inspiral gravitational waves from SMBH binaries with total mass $\gtrsim 10^{9}\, M_{\odot}$ and orbital periods between months and years should lie in the nanohertz (nHz) - microhertz range \citep{ses04,bur18,Burke-Spolaor2019}.
The routine detection of nHz gravitational waves (GWs) from
 SMBH binary systems is the primary goal of rapidly maturing Pulsar Timing Array (PTA) experiments \cite{NG12_5_2020,PPTA21,EPTA21}.
The eventual detection of nHz GW sources will complement the vibrant field of GW astronomy, inaugurated by the routine detections of hectohertz GWs from stellar-mass black hole (BH) binaries by the LIGO-Virgo collaboration \cite{gwtc2,gwtc3}.
This collaboration also heralded the era of multi-messenger GW astronomy by the observations of hectohertz GWs from a binary neutron star merger (GW170817) and its electromagnetic (EM) counterparts (EM170817) \cite{gw170817, gw170817_mm}.
The eventual detection of continuous nHz GWs from individual SMBHBs and observations of their electromagnetic (EM) counterparts can make profound contributions to the emerging field of multi-messenger GW astronomy \cite{Burke-Spolaor2019,xin2021}.
EM observations in recent years have suggested the existence of many candidate binary SMBH systems with varying orbital periods (e.g., OJ~287 with ~12 yr period \citep{sil88}; 3C~66B with ~1 yr period in the radio \cite[][but also see \cite{Jenet2004,Iguchi2010}]{Sudou2003}; 
SDSS~J1201+3003 identified from a  tidal disruption event optical lightcurve \cite{Liu2014}; 
PG1302-102 identified in the optical Catalina transient survey \cite[][see also \cite{Zhu2020}]{Graham2015};
and see, e.g., \citet{charisi2016} who found 33 further quasars with statistically significant periodic variability).  
Some of these systems and candidates are in the nHz GW regime which should be detectable in GWs by the PTA consortium during the SKA era \citep{Zhu2019,dey19,YiFeng20}. 
However, none of these systems had as detailed modeling and testable predictions as
 OJ~287, and no lightcurve has had the very long-term coverage that was obtained for OJ~287.

Persistent multi-wavelength electromagnetic observations will be crucial to pursue multi-messenger GW astronomy with OJ~287 in the coming decades.
In what follows, we provide a brief introduction to the binary black hole (BBH) central engine description for OJ~287 and discuss the context and the need to pursue a variety of electromagnetic observations. 
These efforts should allow us to place tight observational constraints on the various aspects of the BBH central engine description for OJ~287 that will be crucial for pursuing  multi-messenger GW astronomy with OJ~287. 
We now describe briefly our BBH description for OJ~287 and ongoing multi-wavelength observational campaigns on the source.
 

\subsection{OJ~287 and its BH binary central engine description}
\label{subsec:BBH_model}
The bright blazar OJ~287 (RA: 08:54:48.87, Dec: +20:06:30.6), situated at a redshift of $z = 0.306$, was first discovered as a radio source during the course of the Ohio sky survey in the 1960s \citep{dickel67}. 
However, the optical light curve of OJ~287 goes all the way back to the year 1888 as it had been unintentionally photographed often in the past due to its proximity to the ecliptic plane \citep{sil88, Hudec2013}.
Interestingly, the optical light curve of OJ~287, extending over $130$ years, exhibits unique quasi-periodic high-brightness flares (outbursts) with a period of $\sim 12$ years \citep{sil88, val06, Dey18}.
Further, the blazar shows a long-term variation in its apparent magnitude with a period of $\sim 60$ years \citep{val06} (see Figure~\ref{fig:lightcurve-optical}).

In 1982, the binary nature of OJ~287 was recognized by Aimo Sillanp\"a\"a, while constructing historical light curves for the quasars in the Tuorla - Mets\"ahovi variability survey which had begun two years earlier \citep{kid07}. 
It was obvious that the next major outburst was about to happen and the blazar monitoring community was alerted. This first OJ~287 campaign was fruitful \citep{smi85,sil85}. 
A more detailed explanation of the binary BH (BBH) scenario was subsequently proposed to explain the quasi-periodic flares in OJ~287 by associating them with increased accretion flow to the primary BH. 
This was due to the accretion disk perturbations of the secondary BH during its pericenter passages in an orbit which lies in the disk plane \citep{sil88}.
Moreover, a notable prediction of the model was that OJ~287 should show an outburst again in the autumn of 1994. This was verified by the second campaign called OJ-94 \citep{sil96a}.

In 1994, \citet{LV96} recognized that the flares in OJ~287 were not exactly periodic, and that the systematics of the past flares are better understood if the flares are double-peaked and the two peaks are separated by $\sim 1-2$ years.
This led to the proposal of a new SMBHB central engine model (LV96 model hereafter) for OJ~287 where a supermassive secondary BH is orbiting the more massive primary BH in a relativistic eccentric orbit with a redshifted orbital period of $\sim 12$ years and where the orbital plane is inclined with respect to the accretion disk of the primary at a large angle.
In this model, the double-peaked flares occur due to the impacts of the secondary BH on the accretion disk of the primary BH twice during each orbit. The third campaign carried out by the OJ-94 group verified the flare on October 1995. It came within the narrow two-week time window of the prediction \citep{val96,sil96b}.

Each impact releases hot bubbles of gas from both sides of the disk which then expand and cool down until they become optically thin \citep{LV96, iva98}. At this point, bremsstrahlung radiation from the entire volume becomes observable causing the optical brightness of the blazar to increase rapidly \citep{Pih16}.
The resulting flare can last for a few weeks and should reach its peak within the course of a few days.

The LV96 BBH central engine model of OJ~287 with subsequent fine-tuning of the orbital and disk parameters has been very successful in explaining the past observed optical flares, including many which have been discovered in historical data since 1994.
See \citet{Dey18} for a collection of flare light curves. Moreover, this model has accurately predicted the time of arrival of future flares expected during the years 2005, 2007, 2015, and 2019 which triggered new observing campaigns \citep{sun96,sun97,val06,val07, Dey18}. 
The 2005 flare came a week ahead of the orbital schedule calculated by \citet{sun96} and corrected for the astrophysical time shift between the disk impact and the flare by LV96 \citep{val06a,val11b}. The latest prediction of the 2019 \textit{Eddington flare} was successfully observed with the {\it Spitzer space telescope} \citep{Laine20} which showed that the orbital times are now known with the accuracy of $\pm4$ hours.

The model also predicted that the excess emission during the impact flares is unpolarised, and it has the bremsstrahlung spectrum at $\sim~ 3\times10^5$ K. At the time (in the spring of 1995 when the model was constructed) it was only known that during the rapid brightening of the 1983 flare the polarisation decreased with rising flux: the first point at $\sim~16.5$ mJy in the V-band had polarisation of $8.2~ \%$ while the peak at $\sim~33$ mJy had polarisation of $0.8 ~\%$ and the four points measured around the peak had the average polarisation of $4~\%$ \citep{smi85}. This is exactly what is expected if the excess emission is unpolarised. The same trend was seen again in 2007 and 2015 \citep{val08,val16}. In 1984, 1994, 1995, 2005 and 2019 the polarisation measurements were not obtained for the primary peak, partly due to seasonal and weather restrictions.

The spectral slope over a wide range of frequencies was measured during the 2005 flare and the expected spectral index of $\alpha_\nu\sim-0.2$ was found. It is quite different from the usual spectral index of $\alpha_\nu\sim-1.3$ over the same spectral range \citep{val12,kid18}. The spectral index $\alpha_\nu\sim-0.2$ was confirmed during the 2015 and 2019 flares. The complete change in the nature of emission in the flare as compared with the usual out-of-flare emission excludes the possibility that the flares are somehow related to Doppler boosting variations in a turning jet \citep{vil98,rie04}.

As an example, the 2015 flare was observed by the Swift observatory \citep{cip15}. The first observation prior to the flare on JD 2457354 found the flux 1.92 mJy in the UV channel M2. Almost simultaneously it was observed in several optical bands, including the V-band, where the flux was recorded as 5.67 mJy \citep{mac15}. The highest M2 flux value in this set of Swift observations was recorded on JD 2457360.65: it was 5.55 mJy, while the nearly simultaneous V-flux (about 0.5 d earlier) was 10.0 mJy \citep{sha15}. Our R-band magnitude 13.36 on JD 2457360.65 is in agreement with this V-flux.  We take the first observations as the base level, and obtain the flare contributions in M2 and V-bands of 3.63 mJy and 4.33 mJy, respectively. The spectral index between these two wavebands becomes $\alpha_\nu\sim-0.2$, the bremsstrahlung value for the temperature in question.

The observation of the bremsstrahlung optical flare during early September 2007 confirmed the highly-relativistic nature of the binary BH in OJ~287. This is because the BBH model would have predicted the arrival of OJ~287's impact flare on Earth around early October 2007 if the effects of GW emission had been ignored while describing the relativistic trajectory of the secondary BH in the blazar \citep{val07, val08}.
Thereafter, the observation of the \textit{GR centenary flare} during November-December 2015 was used to constrain the dimensionless Kerr parameter of OJ~287's primary BH \cite{val16, Dey18}.
The updated BBH parameters of the OJ~287 system, 
determined using post-Newtonian (PN) equations of motion, are given below: primary BH mass $m_1 = 18.35 \times 10^9 M_{\odot}$, secondary BH mass $m_2 = 150 \times 10^6 M_{\odot}$, primary BH Kerr parameter $\chi_1 = 0.38$, orbital eccentricity $e = 0.657$, and orbital period (redshifted) $P = 12.06$ years \cite{Dey18}.
We now describe the various EM observational campaigns 
that allowed us to extract additional astrophysical details of 
the BBH central engine description for the blazar. 

\subsection{Multi-wavelength emission of OJ 287 and previous and ongoing monitoring campaigns}  

\begin{figure}
\centering
\includegraphics[width=0.6\textwidth]{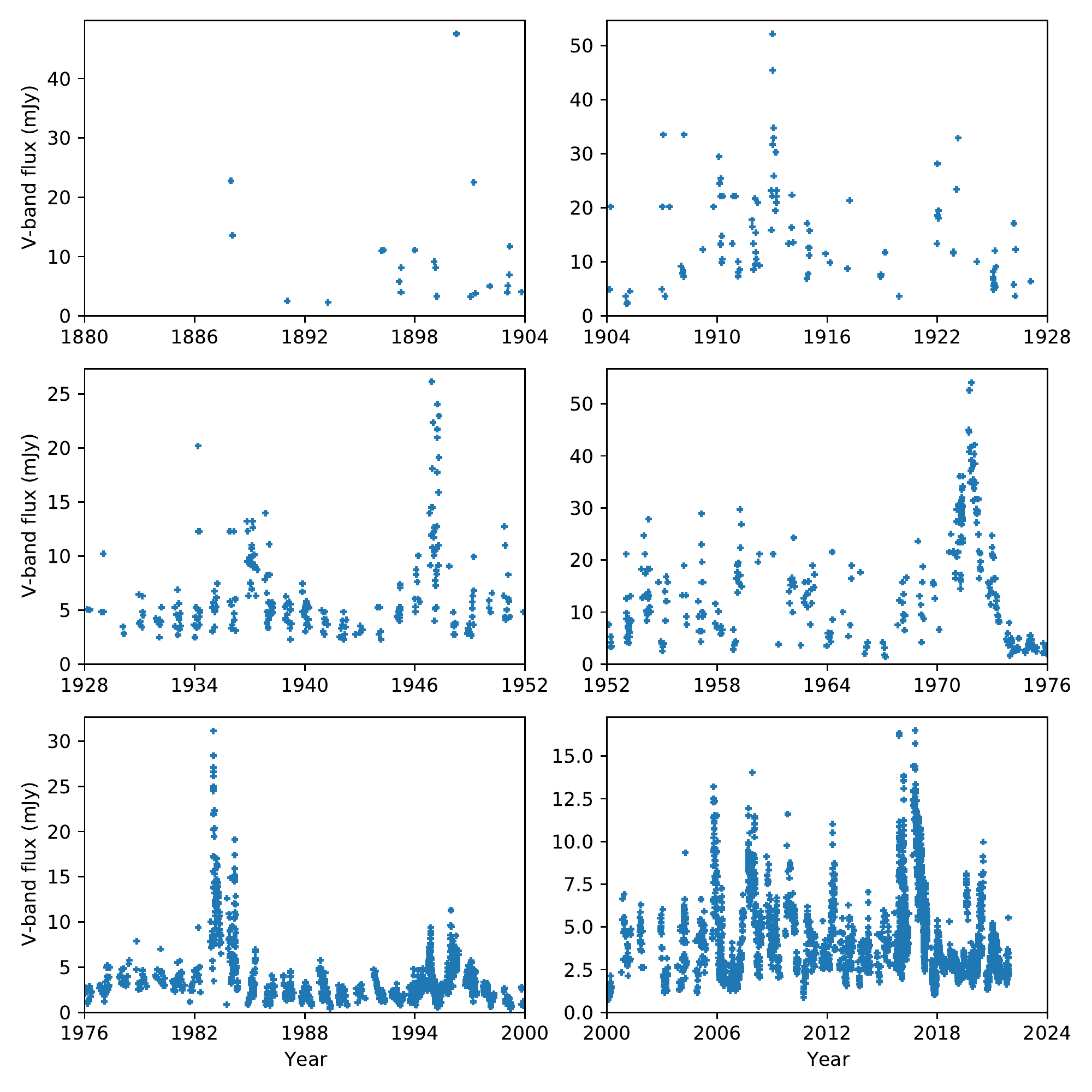}
	\caption{Optical lightcurve of OJ~287 between 1888 and 2021. The figure includes a large number of previously unpublished data points by two of the authors (R.H., S.Z.). All point have been transformed to the V-band, including data in \citet{Laine20}.}
     \label{fig:lightcurve-optical}
\end{figure}

OJ~287 has been the target of intense multi-wavelength coverage
and monitoring in recent decades. 
Early optical observations first revealed very bright high-states reaching up to
$\sim$12 mag. Dedicated optical monitoring campaigns \citep{Pursimo2000,Villforth2010,Pur21} and the addition of photographic plate data dating back to the late 19th century \citep{Hudec2013} then revealed that bright optical maxima recur semi-periodically.
These observations led to first suggestions of the presence of a binary SMBH in OJ~287 \citep{sil88} and its later confirmation in subsequent campaigns, as mentioned above \citep{val11b, val17}. 
An ongoing optical monitoring program of OJ~287 combines optical data taken at multiple ground-based observatories and provides coverage at sub-daily cadence except for epochs when OJ~287 is unobservable due to its proximity to the Sun \citep{Zola2016,Dey18}.
Figure~\ref{fig:lightcurve-optical} shows the optical lightcurve of OJ~287 from late 19th century to the present epoch.
Based on the latest optical high-state discovered in this program in late 2015 and interpreted as a binary impact flare \citep{val16}, a new broad-band radio~--~X-ray monitoring program of OJ~287 was initiated and is described below.  
Most of the time, the optical emission of OJ~287 is dominated by synchrotron emission from the jet,
as known from SED modeling \citep{Abdo2009} and high levels of polarization \citep{DArcangelo2009}.
However, it shows a decrease in polarization, expected in the binary SMBH model, at epochs where an additional thermal bremsstrahlung contribution from the disk impacts contributes to the emission \citep[e.g.][]{val08}. 
Optical polarimetric monitoring of OJ~287 was carried out during the epoch 2016-2017 following changes in polarization and led to the interpretation of the 2016-17 optical high-state as a binary after-flare \citep{val17}.  

Based on the faintness of its optical Balmer lines from the broad-line region \citep[BLR;][]{sit85, nil10}, OJ~287 is classified as a BL Lac object.  Given its spectral-energy distribution (SED),
it is of LBL type \citep[low-frequency-peaked blazar;][]{Padovani1995}. Optical emission lines from the narrow-line region (NLR) like [OIII]$\lambda$5007 are only faintly present \citep{nil10}. 

As a blazar,  OJ~287 is a bright radio emitter \citep{dickel67} with a structured, relativistic jet
\citep{jor05, Lee2020, Gom21} that is highly polarized \citep{Goddi2021}. On average,
the jet is pointing close to our line of sight \citep{jor05, Hodgson2017}, but it shows large swings in its position angle \citep{Agu12, val13} that can be understood in the context of the binary SMBH model \citep{Dey21}.

\begin{figure}
\centering
\includegraphics[clip, trim=1.8cm 2.1cm 2.2cm 0.3cm, angle=-90, width=9cm]{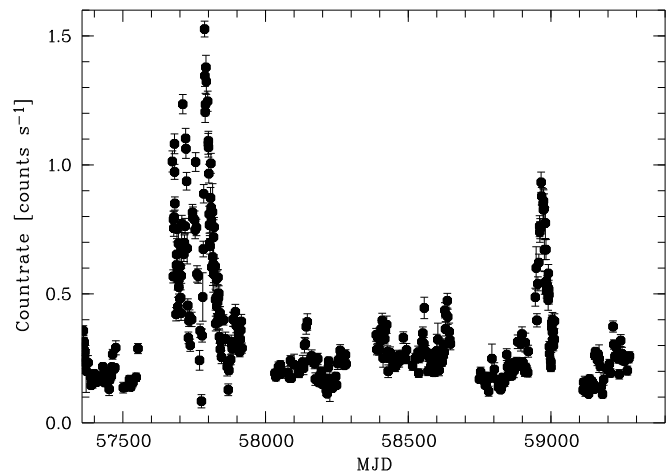}
	\caption{X-ray (0.3--10 keV) lightcurve of OJ~287 obtained in the course of the MOMO program with Swift between 2015 December and 2020 March 1. 
    Two bright outbursts were discovered in 2016-17 and 2020. Adopted from \citet{Komossa2021d}. }
     \label{fig:lightcurve-X}
\end{figure}

OJ~287 was first detected in the $\gamma$-ray regime with the CGRO/EGRET satellite observatory \citep{Shrader1996} and shows repeated flaring in the Fermi $\gamma$-ray band \citep[e.g.,][] {Abdo2009, Agu11, Hodgson2017} along with 
more quiescent epochs not detected by Fermi. The first very-high energy (VHE; $E>100$ GeV) 
detection was obtained in 2017 with VERITAS \citep{Obr17}; an observation triggered by the 
measurement of an exceptionally bright X-ray outburst (Figure \ref{fig:lightcurve-X}) discovered with Swift in the course of
the dedicated long-term, multi-frequency monitoring program MOMO (Multiwavelength Observations and 
Modelling of OJ 287; \citep{Komossa2017, Komossa2020}).  

OJ~287 is a bright and variable X-ray emitter observed with most major X-ray observatories    
and first detected with the Einstein satellite \citep{Madejski1988}.
The X-ray spectrum of OJ~287, based on XMM-Newton 
observations between 2005 and 2020 (energy range: 0.3--10 keV), is characterized by a super-soft 
(low-energy) synchrotron component and a hard (high-energy) inverse Compton (IC) component
\citep{Komossa2021a}. The synchrotron component dominates the X-ray outburst states \citep{Komossa2020}.  High-energy emission up to 70 keV was detected with NuSTAR with an unusually soft power-law 
spectrum of photon index $\Gamma_{\rm X}=2.4$ accompanying the 2020 outburst \citep{Komossa2020}. 
OJ~287 shows an X-ray jet extended by 20" 
\citep{Marscher2011}.  

Starting in late 2015, OJ~287 has been the target of the  dedicated multi-wavelength monitoring program MOMO \citep{Komossa2017, Komossa2020, Komossa2021a, Komossa2021c, Komossa2021d} (see the review by \citet{Komossa2021b}).
MOMO is aimed at characterizing the blazar properties and facets of the binary SMBH in detail. 
The program employs multiple frequencies in the radio band between 2-40 GHz obtained with the 
Effelsberg observatory, UV--optical Swift data in 6 filters, and the Swift X-ray band (0.3--10 keV).  The monitoring cadence is as short as 1-4 days. Additional deeper observations with multiple 
ground and space-based observatories are triggered in the case of exceptional outburst or low-states
discovered with Swift and/or in the radio band. The community is alerted in form of {\sl{Astronomer's Telegrams}} about particular states of OJ~287 (ATel \#8411, \#9629, \#10043, \#12086, \#13658, \#13702, \#13785, and \#14052). 
MOMO is the densest multi-wavelength monitoring program of OJ~287 involving X-rays, with $>$1000 data sets so far taken, providing spectra, broad-band SEDs, and lightcurves, during all activity states of OJ 287 \citep{Komossa2021b, Komossa2021d}. 

\begin{figure}
\centering
\includegraphics[clip, trim=1.8cm 5.6cm 1.3cm 2.6cm, width=13cm]{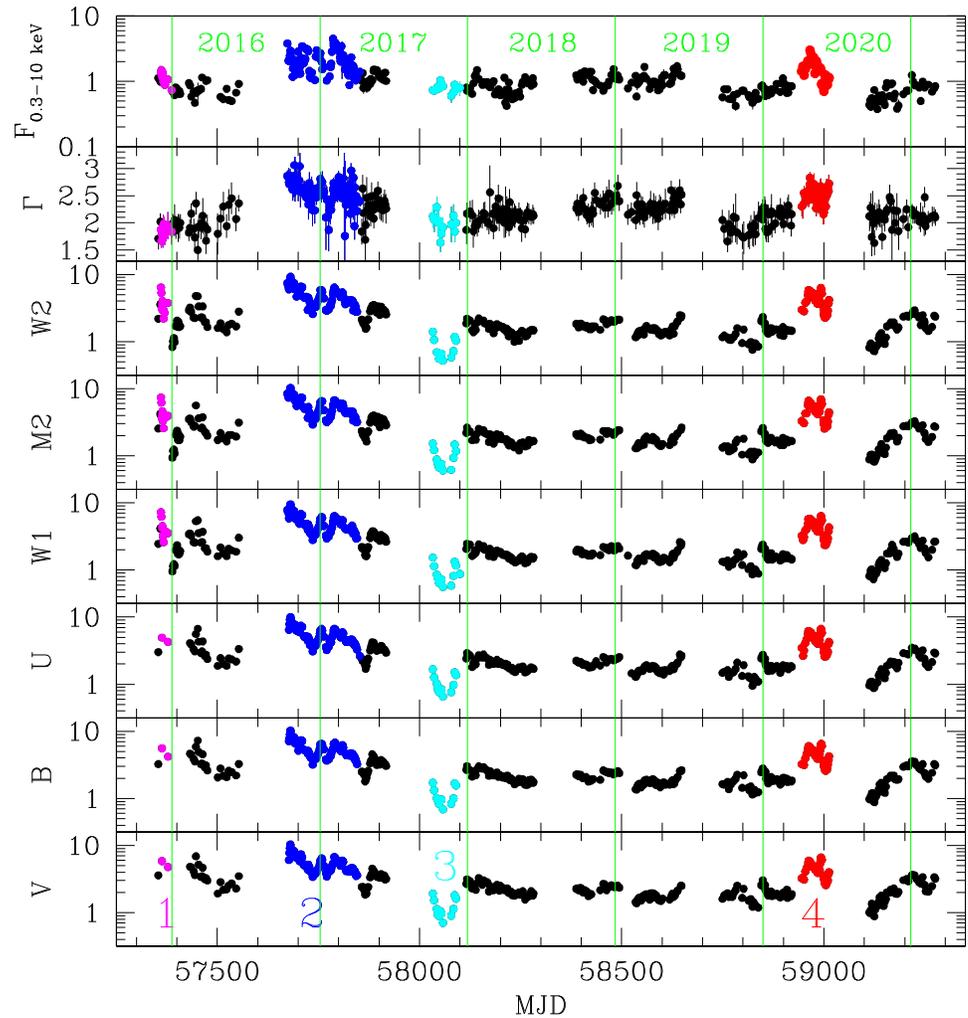}
	\caption{Multi-wavelength X-ray--optical lightcurve of OJ~287 taken with Swift since December 2015. The majority of data was obtained during the MOMO program. Panels from top to bottom: Observed X-ray flux in the band (0.3--10) keV in units of 10$^{-11}$ erg/s/cm$^2$, X-ray powerlaw photon index $\Gamma_{\rm x}$, Swift UVOT UV and optical 
	fluxes (corrected for Galactic extinction) in units of 10$^{-11}$ erg/s/cm$^2$. Every year, observations are interrupted by $\sim$3 months when OJ~287 is in close proximity to the Sun such that Swift cannot observe. Selected epochs are marked in color: (1) The epoch near the 2015 binary impact flare, (2) the bright 2016-17 outburst, (3) the 2017 optical-UV deep fade, and (4) the 2020 April outburst. Epochs (2) and (4) were interpreted as possible binary after-flares.  Epochs like the deep fade in 2017 (feature 3) and the new low-state in September 2020 can be used to obtain host galaxy imaging of OJ 287 when the blazar emission itself is least disturbing. 
	Adopted from \citet{Komossa2021d}. 
     }
     \label{fig:lightcurve-MOMO}
\end{figure}

Some of the main results of the program MOMO are as follows: 
Two bright outbursts were identified in the course of MOMO in 2016-17 and 2020 (Fig. \ref{fig:lightcurve-X}); the brightest so far recorded in X-rays
\citep{Komossa2017, Komossa2020, Komossa2021d}). These outbursts were non-thermal in nature, and they are consistent \citep{Komossa2020} with
after-flares predicted by the binary SMBH model, when disk impacts of the secondary lead to the triggering 
of new jet activity of the primary SMBH. 
OJ~287 has turned out to be one of the most spectrally variable blazars in the X-ray regime 
(Figure~\ref{fig:lightcurve-MOMO}) with power-law spectral indices as steep as $\Gamma_{\rm X}=3$ \citep{Komossa2021a, Komossa2021d}. Supersoft synchrotron emission dominates the outburst states. 
Another key features of the MOMO lightcurve is a pronounced, symmetric  UV--optical deep fade in 2017 November -- December (Figure~\ref{fig:lightcurve-MOMO}) of still unknown nature \citep{Komossa2020, Komossa2021d}. Structure function analyses of Swift data between 2015--2020 has revealed characteristic break times of 4--39 days depending on activity state and waveband  \citep{Komossa2021d}. Discrete correlation analysis between UV, optical and X-ray bands during the same epoch shows lags and leads in X-rays w.r.t. the UV, attributed to an IC component, where external Comptonization is favored over synchrotron-self-Compton emission \citep{Komossa2021b, Komossa2021d}. 

The MOMO program continues as OJ~287 nears its next predicted binary SMBH impact flare. Results obtained so far including $>$1000 broad-band SEDs will serve as a useful data base to identify new activity states of OJ 287 driven by blazar variability in comparison to binary-driven activity, and allow detailed modelling of the broad-band emission from the radio to the high-energy regime at each epoch.  
We now take a closer look at the present description of 
the primary BH accretion disk and propose
a detailed scenario to describe 
the outcome of secondary BH impact with the disk.

\section{Accretion disk in OJ~287}

Our binary model of OJ~287 is based on the interpretation of the huge optical flares as impacts of a secondary black hole on accretion disk of the primary. While the emission level of OJ~287 is between 1 and 2 mJy in the optical V-band most of the time, the emission rises to 13 mJy (in the 2015 flare) or similar values in the peaks that we interpret as impact flares, i.e they represent an order of magnitude increase in the optical flux. The increase occurs on the time scale of a day or so which is about a fraction of 0.0002 of the 12 yr (9 yr redshift corrected) binary orbital period. This implies that the emission region in the source responsible for these huge flux variations, cannot be greater than one light day in size. Most of the work on the binary black hole model has concentrated on the orbit of the secondary. However, it is equally important that we also find out the basic properties of the accretion disk from the way the disk reacts to the impacts, and how these one-light-day emission regions are activated. 

In the original binary black hole model of \citet{LV96} the disk was taken as a standard disk with two parameters, viscosity $\alpha$ and accretion rate $\dot{m}$. Fixed values were given for these parameters. When more impacts were observed later, both in the future and in the past historical light curve, it became possible to solve the values of these parameters together with the orbit solution. Altogether it is a 9 parameter problem which has a unique solution as soon as the exact times of 10 flares are known. Mathematically it is equivalent to a ninth order equation which has nine unique roots. The method of solution is iterative, similar to the Newton-Raphson algorithm \citep{val07}. From the detailed study of the 2015 flare, the values of the two standard parameters were determined to be $\alpha = 0.26$ and $\dot{m} = 0.08$ in Eddington units.

In the model of \citet{LV96}, a spherical bubble of hot plasma is released from the disk by the impact. \citet{iva98} later demonstrated that actually two bubbles are released, one on each side of the disk. Therefore each impact is detectable via a radiating bubble on the front side of the disk. The model further specified that the bubble expands isotropically and therefore retains its spherical shape. The simulation of \citet{iva98} casts doubt on whether this is true at an early stage of the bubble, but this was not considered important, as the bubble becomes optically thin only at a later time which results in the observed great increase in brightness. 

\citet{val19} noted that the maximum brightness of the 2015 flare should have been 26 mJy in the standard model. The fact that the peak value was only half of this value is most easily explained if the bubble is compressed along the line of sight. Thus it appears necessary that the expansion of the bubble is not completely unconstrained, but that it is influenced possibly by the general magnetic fields in the disk.
In that case, it would be interesting to know what happens to the flux of the bubble from the very beginning if its expansion is constrained all the way. From \citet{iva98} one could conclude that initially the expansion is more jet-like, a flow perpendicular to the disk rather than a gentle release of bubbles. \citet{iva98} called them fountains. In the following we consider what difference this makes to the impact flare light curve.

\section{Anisotropically expanding BH impact generated bubbles}

According to Lehto and Valtonen (LV in the following, \citep{LV96}), the compressed plasma inside the accretion disk, following the impact, bursts out of the disk with the speed $v_{perp} = (6 / 7) v_{sec}$ (LV Eq. 7) and the time to come out of the disk is $\tau_{dyn} = h / v_{perp}$ (LV Eq. 8), where $v_{sec}$ is the speed of the secondary black hole at the impact and $h$ is the semi-height of the disk. 
The flux from this bubble follows $S \sim \epsilon^{ff} V_{bubble}$  (LV Eq. 14), if the bubble is fully transparent. 
The quantities $\epsilon^{ff}$ and $V_{bubble}$ are the free-free emissivity and the volume of the bubble, respectively. 
However, we see only the fraction $1 / \tau$ of the bubble, where $\tau$ is  its optical depth. Here $\tau =  d \epsilon^{ff} T^{-1}$, if $d$ is the geometrical depth and $T$ the temperature of the bubble. Therefore the flux goes as
\begin{equation}
S_{seen} \sim \epsilon^{ff} V_{bubble} / ( d \epsilon^{ff} T^{-1}) \,.
\end{equation}
We can substitute $T \sim V_{bubble}^{-1/3}$ in adiabatic expansion which means that
\begin{equation}
S_{seen} \sim V_{bubble}^{2/3} / d \,.
\end{equation}
We can put $V_{bubble} \sim r^2 d$, where $r$ is the bubble radius ($d \sim r$ in spherical geometry). The adiabatic expansion speed $v \sim ~ T^{1/2} \sim ~ V_{bubble}^{-1/6}$ and 
\begin{equation}
d \sim v t \sim V_{bubble}^{-1/6}\, t \,.
\end{equation}                                   
Now we have to specify the geometry of the bubble:
For a tube of fixed radius seen end-on, $V_{bubble} \sim d r^2$, $r = constant$, or $S_{seen} \sim  d^{-1/3}$, and since $d \sim d^{-1/6} t$, it follows that $d\sim ~ t^{6/7}$. Thus
\begin{equation}
S_{seen} \sim t^{-2/7}\,.
\end{equation}                          
Flux drops with time roughly as $S_{seen} \sim t^{-1/3}$.

For a tube of fixed depth seen end-on, $V_{bubble} \sim d r^{2}$, $d = constant$ or  $S_{seen} \sim  r^{4/3}$, and since $r \sim r^{-1/3} t$, it follows that $r \sim t^{4/3}$ . Thus
\begin{equation}
 S_{seen} \sim t^{16/9}.
 \end{equation}                                     
Flux increases with time nearly as $S_{seen} \sim t^{2}$ (c.f. Figure 2 in \citep{val19})

\begin{figure}
    \centering
    \includegraphics[width=0.5\textwidth]{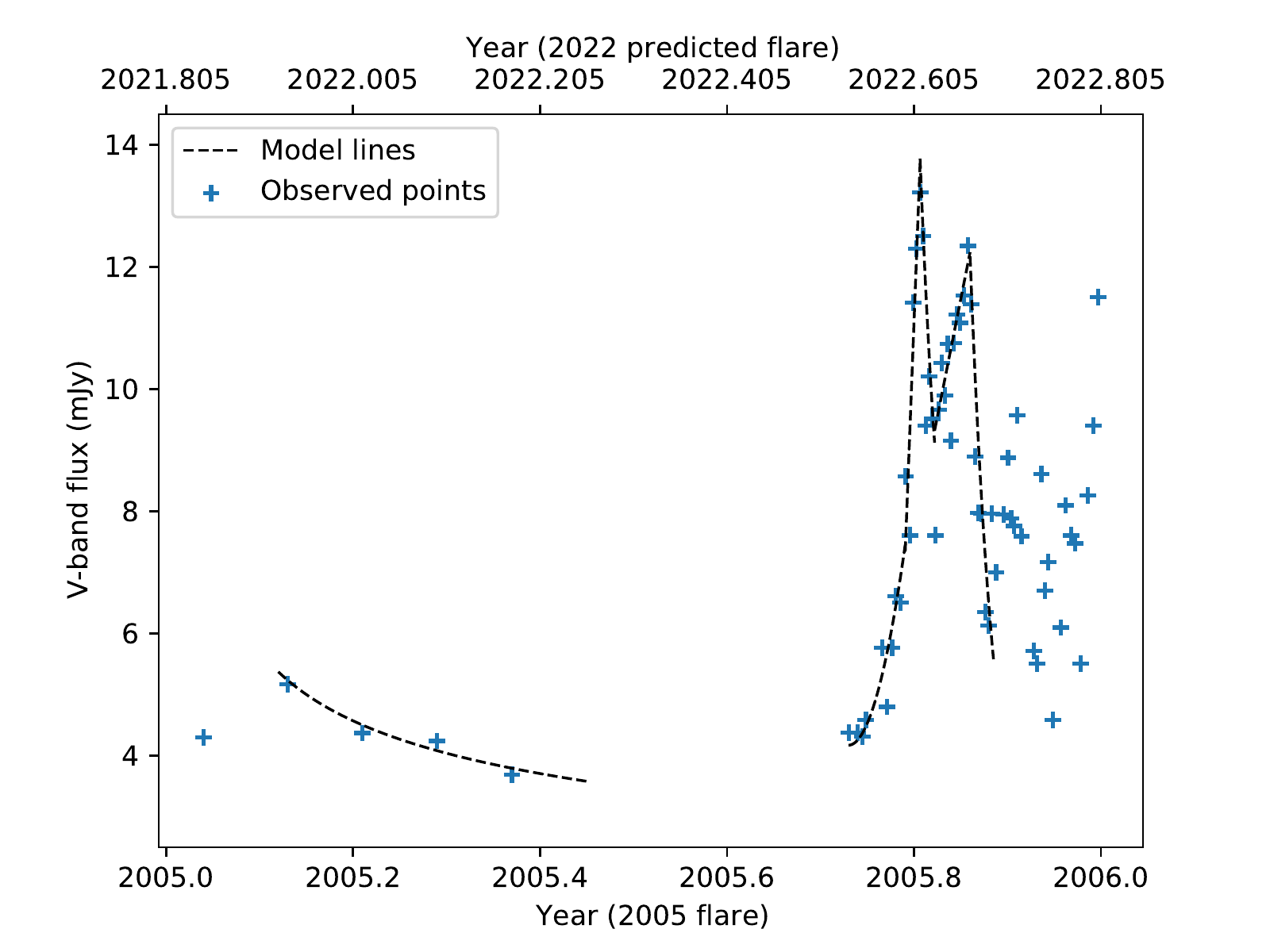}
    \caption{The optical light curve of OJ~287 during  2005 in V-band mJy units. 
    Plot shows the monthly averaged data points 
    for the early part of 2005 while the latter part, with faster variability, has 3 day averaged data points. The dashed curve is the theoretical line based on present work and  \citet{val19}. Additionally, the theoretical curve
    provides the expected light-curve of OJ~287 during 2021-2022
    and we mark its epochs on the upper x-axis 
    while employing the 
     time shift of the SMBH binary central engine description \citet{Dey18}. 
     This is motivated by the fact that the secondary BH 
      impact distances from the center should be roughly the same during 2005 and 2021.
       The first big flare during 2022 is not expected to be visible from any Earth-based facility  while the secondary peak may become barely visible by early September 2022.}
    \label{fig:2005_lc_fit}
\end{figure}

For standard spherical bubbles of radius $r$ the corresponding relation is \citep{Pih16}
\begin{equation}
S_{seen}\sim V_{bubble}^{2/3} / r \sim r \sim t^{2/3}.
\end{equation}   
      
Table~\ref{tab1} lists the disk impacts between 1971 and 2031. The first two columns come from \citep{Dey18}, the numbers from the third column onwards 
are derived in the next section.

\begin{specialtable}
\centering
\caption{Times of disk impacts and bremsstrahlung flares as well as the radii of the $H_{\alpha}$ bubbles, impact speed on the disk, the delay of the $H_{\alpha}$ flare and the time of the $H_{\alpha}$ flare\label{tab1}}
\begin{tabular}{cccccc}
\toprule
\textbf{Impact time}	& \textbf{time of flare}  & {${R_{H{\alpha}}}$}  \textbf{(AU)}   & {${v_{rel}/c}$}  & \textbf{delay (yr)}  & \textbf{${H{\alpha}}$ flare}\\
\midrule

1971.11		& 1971.13       	& 2860     & 0.34   & 0.72    & 1971.83\\
1972.71		& 1972.93           & 3930     & 0.20   & 1.66    & 1974.37\\
1982.83     & 1982.96           & 3160     & 0.30   & 0.81    & 1983.74\\
1984.07     & 1984.12           & 3250     & 0.28   & 0.91    & 1984.98\\
1994.48     & 1994.59           & 3610     & 0.23   & 1.31    & 1995.79\\
1995.82     & 1995.84           & 2940     & 0.33   & 0.76    & 1996.58\\
2005.11     & 2005.74           & 4360     & 0.16   & 2.31    & 2007.42\\
2007.68     & 2007.69           & 2740     & 0.36   & 0.65    & 2008.33\\
2013.48     & 2015.87           & 5000     & 0.12   & 3.52    & 2016.80\\
2019.56     & 2019.57           & 2740     & 0.36   & 0.65    & 2021.21\\
2021.92     & 2022.55           & 4360     & 0.16   & 2.31    & 2024.23\\
2031.40     & 2031.41           & 2900     & 0.33   & 0.74    & 2032.14\\
\bottomrule
\end{tabular}
\end{specialtable}

In the spring of 2005 there was a major flare, 6.6 mJy at 2005.12, one of the biggest deviations from the standard binary black hole model during years 1996-2010 \citep{sun96,sun97,val11a}. Figure~\ref{fig:2005_lc_fit} shows the observations (monthly averages) from 2005, prior to the main flare in October of that year. The disk impact occurred at 2005.11 (February 9, 2005). With a time translation such that December 3, 2021 corresponds to February 9, 2005, we generate a predicted light curve for the 2021/22 season. For the values appropriate for 2005 and 2021 impacts, which take place at the same radial distance from the center,  $\tau_{dyn} \sim 20 $ days. The flaring related directly to the disk impact should start around December 23, 2021, and the flares should be recognizable as bremsstrahlung flares in the optical - UV region, with no counterpart in radio nor X-rays. The on-going and planned observational campaigns should be able to provide excellent constraints on our detailed 
modeling. So far, the period of early December 2021 was covered with Swift in the ongoing MOMO project and with ground-based optical observations. The optical--UV low-state detected in December 2--14 implies a small shift in the onset of the flare (work in preparation). 

We now move on to explore other observational  implications of the expanding impact bubble.
This requires an extension of LV96 \citep{LV96} 
where 
the brightness evolution of the bubbles was terminated at the epoch when the bremsstrahlung flares emerge. However, at a later stage when the bubbles have cooled enough, they may become visible again as hydrogen line flares.
In what follows, we study the timing of these proposed $H\alpha$ flares, and connect them with one epoch when the hydrogen lines were seen to be exceptionally bright, by a factor of ten above the normal.

\section{Possible spectroscopic implications of the evolving BH impact generated bubbles}

After the bremsstrahlung flare the radiating bubble cools until it falls below the jet continuum level and is not observed any more \citep{val19}. However, at a later time the bubble has cooled enough to become an $H II$ region. This may appear as a sudden increase of the strength of the Balmer lines for a limited period of time. At the end of 1984, \citet{sit85} observed such an increased line flux, approximately a factor of ten higher than normal \citep{nil10}. To our knowledge, this was a unique occasion when such an $H\alpha$ flare has been seen. Therefore it is tempting to associate this epoch with the dying moments of one the bubbles that had generated a bremsstrahlung flare at an earlier time.

In order to estimate when the bubbles left over from disk impact may reach the $H\alpha$ stage, we calculate the required increase in the bubble radius until its temperature has dropped to $10^{4}$ K. Using Table 3 of \citet{Dey18} and Table 1 of \citet{val19}, and assuming an adiabatically expanding bubble, we obtain the radii of the bubbles listed in column 3 of Table~\ref{tab1}. In \citet{val19} the expansion speed of the bubble is given as $\sim v_{rel}$/4 where $v_{rel}$ is the speed at which the secondary impacts the accretion disk. In fact, the expansion speed value $v_{rel}$/4.6 is used in the following since it gives the correct timing for the 1984 $H\alpha$ flare, 1984.98 \citep{sit85}. This value is well within the uncertainty of the expansion speed, extracted from the numerical simulations by \citet{iva98}. These values are given in column 4 of Table~\ref{tab1}, while column 5 gives the time of the expansion up to this stage. In column 6 we give the estimated time of the beginning of the $H\alpha$ flare.

The epochs of observation in the line monitoring program of \citet{nil10} and in the program by some of the present authors at the {\it Nordic Optical Telescope} (T.P.) as well at the {\it Rozhen Telescope} of Bulgarian Academy of Sciences (S.Z.) are given in Table~\ref{tab2}. 

\begin{specialtable} 
\centering
\caption{The epochs of the line monitoring program.\label{tab2}}
\begin{tabular}{cccc}
\toprule
\textbf{time}	& \textbf{telescope}	& \textbf{Strong lines?}  & \textbf{secondary line}\\
\midrule

2005.26		& VLT			& no      & (8059)\\
2005.89		& VLT			& no      & (8092)\\
2006.25		& VLT			& no      & (8059)\\
2006.93		& VLT			& no      & (8637)\\
2007.26		& VLT			& no      & (9405)\\
2008.01		& VLT			& no      & [7418]\\
2008.27		& VLT			& no      & [7787]\\
2010.12		& NOT			& no      & 6438\\
2010.81		& NOT			& no      & 6494\\
2010.88		& NOT			& no      & 6499\\
2011.01		& NOT			& no      & 6506\\
2011.05		& NOT			& no      & 6508\\
2011.11		& NOT			& no      & 6512\\
2011.31		& NOT			& no      & 6521\\
2011.35		& NOT			& no      & 6522\\
2011.75		& NOT			& no      & 6538\\
2011.88		& NOT			& no      & 6542\\
2012.14		& NOT			& no      & 6549\\
2012.99		& NOT			& no      & 6564\\
2013.18		& NOT			& no      & 6566\\
2015.94		& Rozhen			& no   & (8273)\\
2016.26		& Rozhen			& no    & (8300)\\
2021.82		& NOT			& no     & 6708\\
\bottomrule
\end{tabular}
\end{specialtable}

 The expected flares in the time span and close to these observations were at 2007.42, 2008.33, 2016.80 and 2021.21. In the first two cases the observational epochs were too early by 0.16 and 0.05 yrs, respectively, in order to catch the beginning of the flares. The last two were too early by 0.54 yr and too late by 0.61 yr. Therefore, if the $H\alpha$ flares started as expected, we would not have seen them, even if the flares lasted several months.

In order to calculate the line flux, we proceed as follows. The emissivity of the Balmer $H{\beta}$ line may be taken as 
 \begin{equation}
\epsilon_{H{\beta}} = 8.3\times10^{-26} ~n^2 ~ erg ~cm^{-3}~ s^{-1}\,,
\end{equation}   
where $n$ is the number density of Hydrogen atoms. We calculate the line flux using the density for the 1984 line flare, $n = 3.2\times10^8$ and the bubble radius 3250 AU from Table~\ref{tab1}. The result is
\begin{equation}
 S_{H\beta} = 2.2\times10^{-15}~ erg~ s^{-1}.
\end{equation}
\citet{sit85} find $S_{H{\beta}} \sim 1.2\times10^{-15}~ erg ~s^{-1}$, but state the value as uncertain. The more certain $S_{H{\alpha}} = 24\times10^{-15}~ erg~ s^{-1}$, divided by some number of order of 10 \citep{sit85} takes us close to the above estimate. The radius and the number density are different in 2024 than in 1984, but the different factors practically cancel each other. Therefore we expect the 1984 $H\alpha$ line flare to repeat itself in 2024.

\section{Explaining observed 
gamma ray flares}
Here we will distinguish the classical gamma-ray regime, as observed by the Fermi satellite, with many gamma flares detected (typically in the $E \sim 100$ MeV regime); and the VHE emission ($E > 100$ GeV) that this Section mostly focuses on.  At VHE, OJ~287 was only detected once, by VERITAS. The VERITAS observation was triggered by the Swift detection of an exceptional X-ray high-state in the course of the MOMO program \citep{Komossa2017, Komossa2020}. Otherwise, OJ~287 is monitored in gamma-rays with Fermi, with many large and small flares.

Classical gamma ray flares have been reported many times in OJ~287 \citep{Agu11,Obr17}. For example, the 2009 classical flare coincides with the last of the predicted tidal flares following the 2007 bremsstrahlung flare \citep{sun97} while the 2017 VHE flare came soon after the last of the predicted tidal flares following the 2015 bremsstrahlung flare \citep{Pih13}. We can predict the next VHE gamma ray flare following the same principles. 

\begin{figure}
\centering
\includegraphics[width=0.5\textwidth]{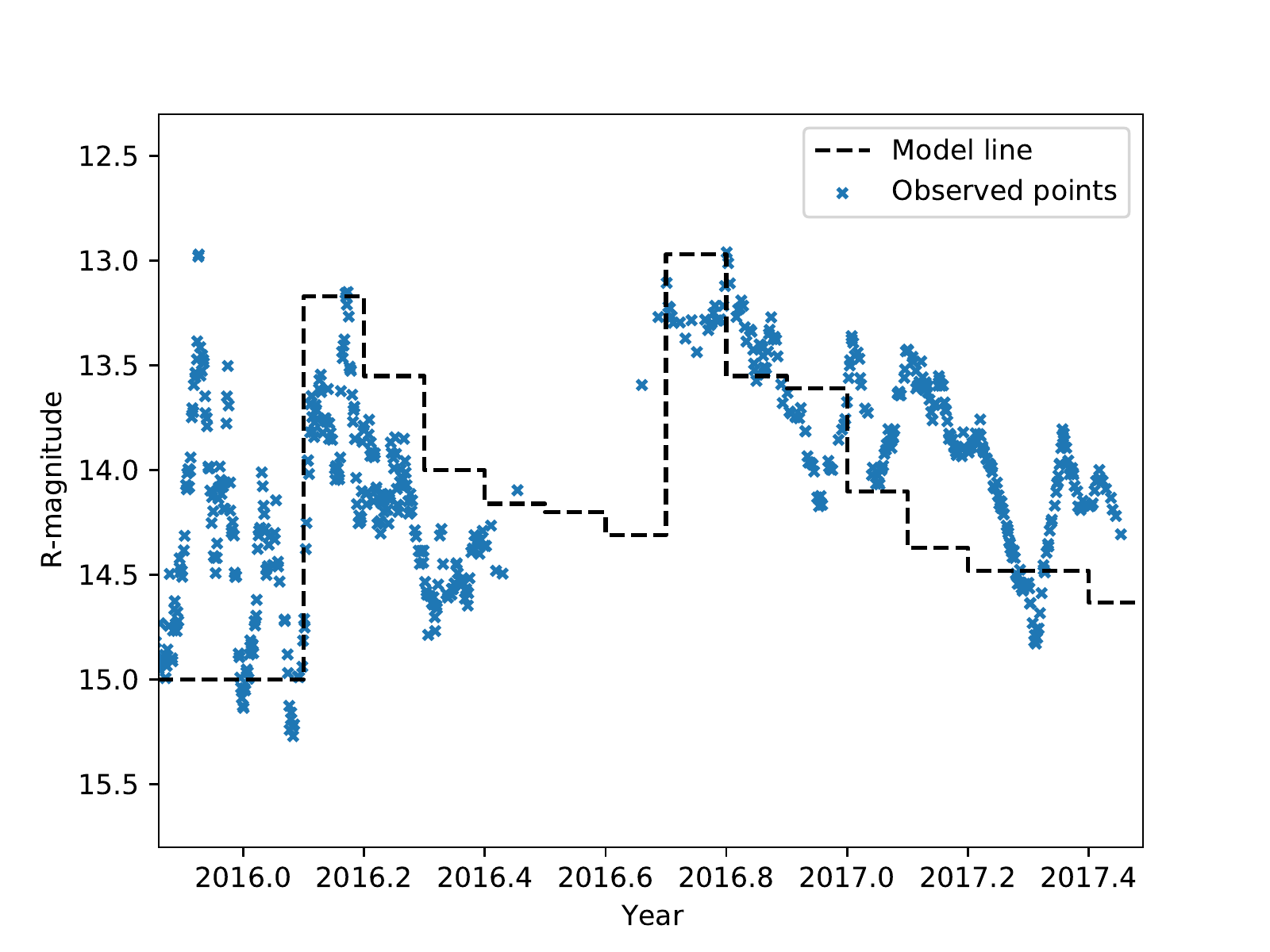}
\caption{Simulation of the response of the accretion disk to the crossing of the secondary black hole through the disk before the 2015 bremsstrahlung flare. Particles which escape the disk are counted, and they are placed on the time axis assuming that the particle contributes to the light curve after traveling the distance of $\sim 15,000$ AU with a constant speed of $0.22 c$ where $c$ is the speed of light. The exact value of travel distance depends on the radial distance from the center where the particle is released from the disk. The particle count in each time box ($time2$) is turned into a magnitude, assuming that the relation between the particle number and the total brightness in each box is linear. The first flare in observations at the end of 2015 is different (impact flare) and is not modeled by this process.\label{fig:2016-2017_lc}}
\end{figure}

In order to study the orbital phase of the 2017 VHE flare in our model, we have calculated the response of the OJ~287 accretion disk to the 2013 secondary crossing, using a disk of 146880 particles. The particles leaving the disk are counted, and the time ($time1$) and distance ($r$) from the center are noted down. Assuming a constant transit speed to the jet (taken to be at $r=3000$ AU), the jet reaction time ($time2$) is calculated. Figure~\ref{fig:2016-2017_lc} corresponds to the transit speed of $c/6$ in the observers frame which equals to $0.22 c$ in the OJ~287 frame. This transit speed value fits best with the two observed light curve peaks at 2016.2 and 2016.8. The present calculation differs from \citet{Pih13} in that the surface density of the disk follows \citet{LV96} rather than a uniform density which was used in \citet{Pih13}. Also \citet{Pih13} did not take the transit time into account, so that it would have to be done collectively for all particles for a comparison with observations. Note that the second peak becomes stronger when a realistic surface density model is used, in agreement with observations.

The tidal flares of the light curve of OJ~287 following the 2005 and 2022 bremsstrahlung flares were calculated by \citet{Pih13}. No time transformation was applied so that the plots in Figure~4 of \citet{Pih13} are for $time1$, as defined above. Using the same transit speed of particles to the jet as above we should add 1.0 yr to the time axis. For the 2005 impact the two main flares in the predicted light curve are at 2007.4 and 2008.6. Even though these are placed at the summer period of few or no observations (OJ~287 is close the sun at the beginning of August), a strong brightening of the source was observed in the following autumn (both 2007 and 2008) which agrees with the tidal flare explanation. The interpretation of the light curve is made complicated by the fact that the bremsstrahlung flare of 2007.7 falls in the same time interval as the tidal flare \citep{val11b}.

As to the 2022 case, the predicted light curve has maxima at 2023.5 and 2024.2 \citep{Pih13}. Note that the nature of the simulation by \citet{Pih13} is such that the tidal flares appear to get weaker with time. This is an artifact of the simulation technique which does not replace the escaped particles by new ones in the disk. Because of practically the same impact distance in 2005 and 2022, the tidal responses should be rather similar. We therefore expect that VHE gamma rays would again be detectable in the spring of 2024. The prediction cannot be more precise until we see the corresponding flare starting in the optical region. 
Further, we expect to see such flare again in the X-ray regime, as in 2017 when the very bright X-ray flare detected with Swift triggered the VHE observation.
Given the ongoing monitoring program with Swift with very dense coverage and $>$1000 SEDs already obtained \citep[e.g.,][]{Komossa2020,Komossa2021d}, very detailed timing and comparison with the 2017 flare will then be possible. This ongoing project, or any X-ray monitoring with future X-ray missions, can then also be used to trigger new VHE observations. 

In addition, OJ~287 observations 
can reveal interesting astrophysical information
when it enters low luminosity state and this is what we focus on in the next section.

\section{Optical multicolor photometry at minimum light}

\begin{figure}
\centering
\includegraphics[width=0.48 \textwidth] {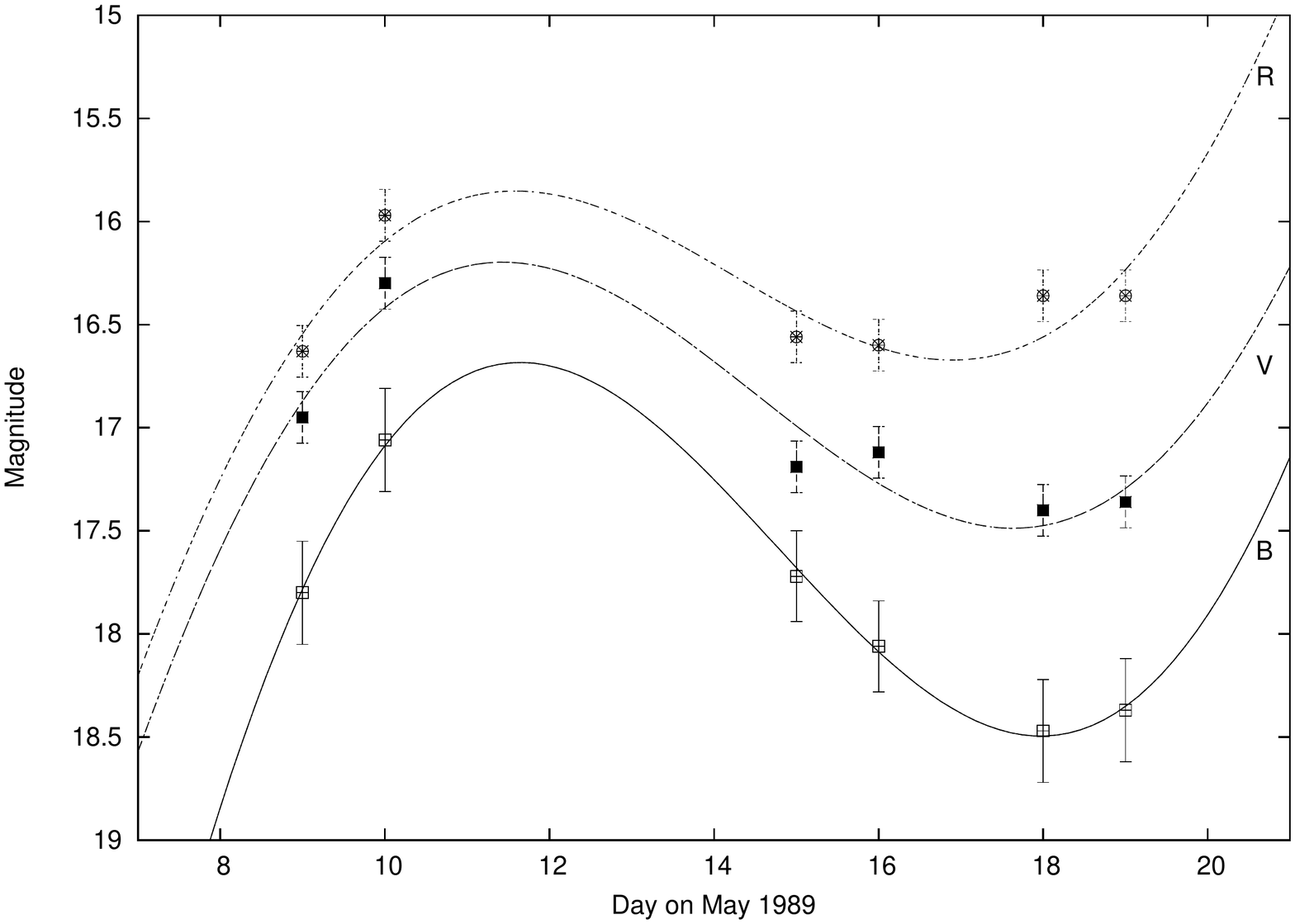}
\caption{The observations of the 1989 fade of OJ~287 with the {\it Nordic Optical Telescope} and the {\it Jacobus Kapteyn Telescope} at La Palma, with $2 \sigma$ errorbars. Here we plot the magnitudes in three filters B, V and R: the I band data were too few to show a significant variation. The lines are fifth order polynomial fits through the data points. We note that the dip on May 18 was deepest in the B band and less deep going to longer wavelengths. There is a possible time delay in the sense that the dip starts from longer wavelengths and happens later at shorter wavelengths. At 37 GHz the minimum took place on May 3, 1989}
\label{fig:1989_fade}
\end{figure}

From time to time, the brightness of OJ~287 falls to a very low light level. It is hard to predict when these times will happen again, even though it is thought that the low light level may be associated with the secondary moving in front of the primary \citep{tak90}. The absolute minimum light observed so far took place in the spring of 1989. In Figure~\ref{fig:1989_fade} we re-plot the observations \citep{tak90}, and fit smooth curves through them. When the brightness of the AGN falls towards the brightness of the host galaxy, the color of OJ~287 is expected to change gradually from the typical jet color to the color of the host galaxy as the fade proceeds. \citet{tak90} conclude that at the brightness level V = 17.4, the host galaxy contributes at least half of the brightness of the source. Using the usual conversion V $-$ K = 3.2, K $-$ correction and evolution correction of 0.6 and the distance modulus m - M = 41, this implies the absolute K-band magnitude of $M_{K} = -26.8\pm 0.3$ \citep{val21}.

Using  surface photometry in the K-band with the {\it Nordic Optical Telescope} \citet{nil20} find $K=-14.65\pm0.6$, leading to $M_{K}=-25.85\pm0.6$ with the K-correction -0.5, which is somewhat, but not significantly, fainter than the value measured by the color method. However, this may not be the total brightness of the host. It is well known that galaxies of this brightness category have extensive haloes which contribute as much as one magnitude to the host. For example, the absolute K-magnitude of NGC~4889 is $M_{K}=-25.4$ \citep{san72} (using m $-$ M = 34.7 and V $-$ K = 3.2) measured out to roughly the same 30 kpc linear distance as the surface photometry of OJ~287 in \citet{nil20} while its total magnitude out to 100 kpc distance is about $M_{K}=-26.8$ \citep{gra13}. Typically the outer envelope contributes about 0.8 magnitudes to the total brightness \citep{dev70,las14,hua13}. Actually, if taken to the redshift of $z = 0.3$, NGC~4889 would probably appear very much like the OJ~287 host. Even the estimated mass of its central black hole is practically the same as the primary BH mass in OJ~287 \citep{gra13}.

We also note that the color photometry method is more likely to capture the faint halo light than the surface photometry method since the latter is seriously influenced by the high background light level generated by the fringes of the strong point source in the center \citep{yan97}. Also the signal-to-noise ratio is better at longer wavelengths than in the traditional optical range which may be the reason why it is difficult to get a host magnitude measurement in the shorter wavelength bands below the infrared K-band \citep{nil20}. For example, the \textit{Hubble Space Telescope} image gives the range from $M_{K}=-24.7$ to $M_{K}=-26.3$, converted from the I-band to the K-band in the usual way \citep{yan97}. The magnitude of the i-band image taken at {\it Gran Telescopio Canarias} corresponds to $M_{K}=-25.2\pm0.3$ \citep{nil20}.

\begin{figure}
\centering
\includegraphics[width=0.48\textwidth] {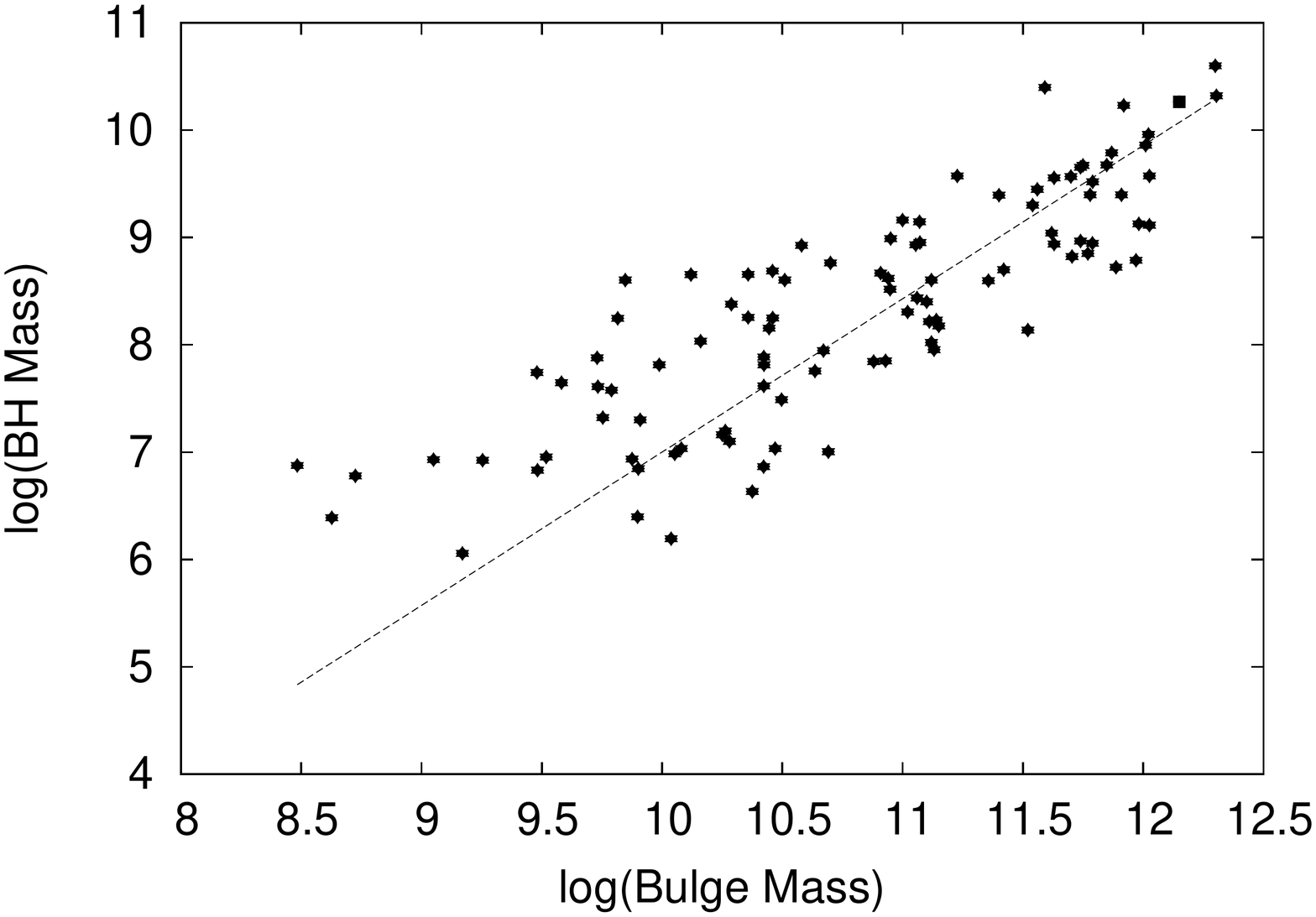}

\caption{The galaxy sample of \citet{sag16} with three additional galaxies in the black hole mass vs. galaxy stellar mass diagram (stars), while the position of OJ~287 is given as a solid square. The dashed line represents the correlation found by \citet{por12} for galaxies in the upper - right hand part of the graph around redshift 0.5.}
\label{fig:BH_mass_host_galaxy}
\end{figure}

It is interesting to see where this magnitude by multicolor method places OJ~287 in the black hole mass - host galaxy mass diagram. Considering that OJ~287 is at a much higher redshift than galaxies normally displayed in this diagram \citep{sag16} and that the correlation is redshift dependent \citep{por12}, we find that OJ~287 follows almost exactly the well established correlation (Figure~\ref{fig:BH_mass_host_galaxy}).

\begin{figure}
    \centering
    \includegraphics[width=0.6\textwidth]{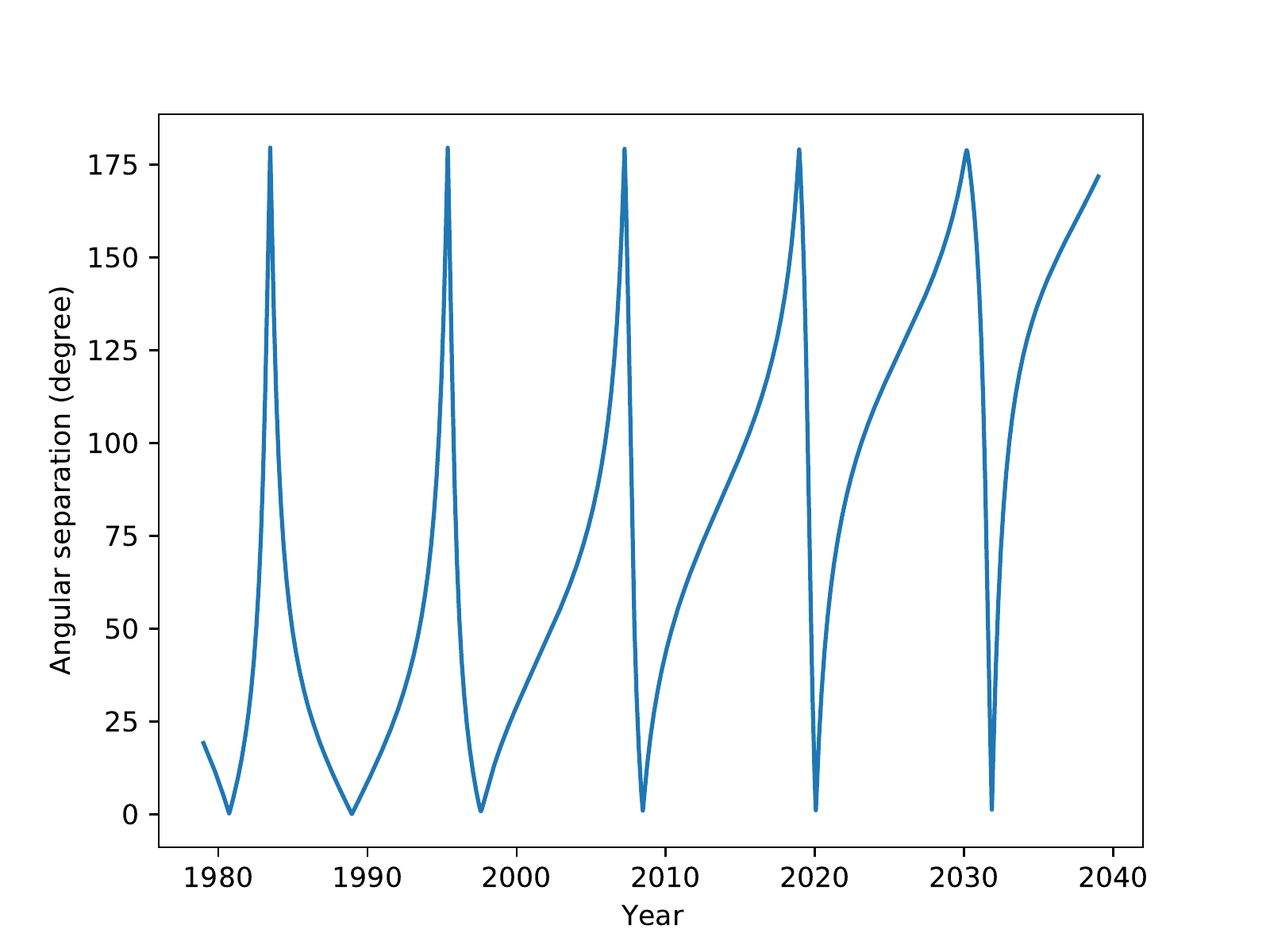}
    \caption{Temporal evolution of the angular separation between the primary jet in OJ~287 and the line joining the primary and secondary BHs. The separation is calculated assuming that the jet precession follows the disk model of \citet{Dey21}. Minimum separation angle occurs at 1980.73, 1988.95, 1997.61, 2008.45, 2020.05, and 2031.85.}
    \label{fig:ang_sep_sec_BH}
\end{figure}

When do we expect to be able to repeat the host galaxy brightness measurement by the photometric method? It depends on the reason behind the fade. At the time in 1989 the secondary was near apocentre of its orbit, and the major axis of the orbit was close to perpendicular to the disk. This configuration minimizes the tidal perturbation of the disk by secondary. The next time a similar configuration arises is around 2032 (see Figure~\ref{fig:ang_sep_sec_BH}).

\citet{tak90} suggests that during the fade the secondary moves between us and the primary, and this causes a temporary misalignment of the jet, leading to decreased Doppler boosting that shows up as a fade. This is a sensible proposition as we may find as follows: Suppose the secondary passes within 100 AU of the jet. Then the jet is gravitationally deflected by $\sim 4^{\circ}$. Using the parameters from \citet{hov09}, and adding $4^{\circ}$ to the viewing angle, we find that the jet becomes fainter by about 3 magnitudes. That is, if we have the normal V-band magnitude $\sim15$, in the fade it may drop to $\sim18$ which is actually what happened during the 1989 fade. We cannot perform a more detailed calculation since we do not know the inclination of the orbit of the secondary. This proposition could be studied further, as the jet swinging back and forth should appear also in radio maps, with the proper delay for each frequency, as well as in optical polarisation with a small delay. A comparison of optical and radio light curves at the fades should also establish the relative positions of the radio and optical cores in the jet. The 1989 fade suggests that they lie close to each other. 

We may ask when this configuration repeats itself. Figure~\ref{fig:ang_sep_sec_BH} and its caption lists such times. Since the optical emission region is in the jet a little way from the center, the times given by the model are about half-a-year earlier than the response in the optical emission. The 1998 and 2008 fades were discussed in \citet{pie99} and \citet{val11a} while the 2020.05 alignment, with the same response time as in 1989, would place the fade in the summer period of 2020 when we had no observations (Figure \ref{fig:ang_sep_sec_BH}).

Obviously a fade would be a good time to repeat the surface photometry study of the host, too. In addition, epochs of very faint intrinsic emission from this blazar (jet) can be used to trigger host galaxy imaging because the central point source and its point spread function (PSF) is least affecting the host measurements at such epochs. For instance, in November 2017 and September 2020 such optical--UV low-states were identified in Swift monitoring (Fig. \ref{fig:lightcurve-MOMO}) and new ones will be detected in the course of the ongoing MOMO program as well as in ground-based optical monitoring campaigns.

\section{The radio jet}

\begin{figure}
    \centering
    \includegraphics[width=0.55\textwidth]{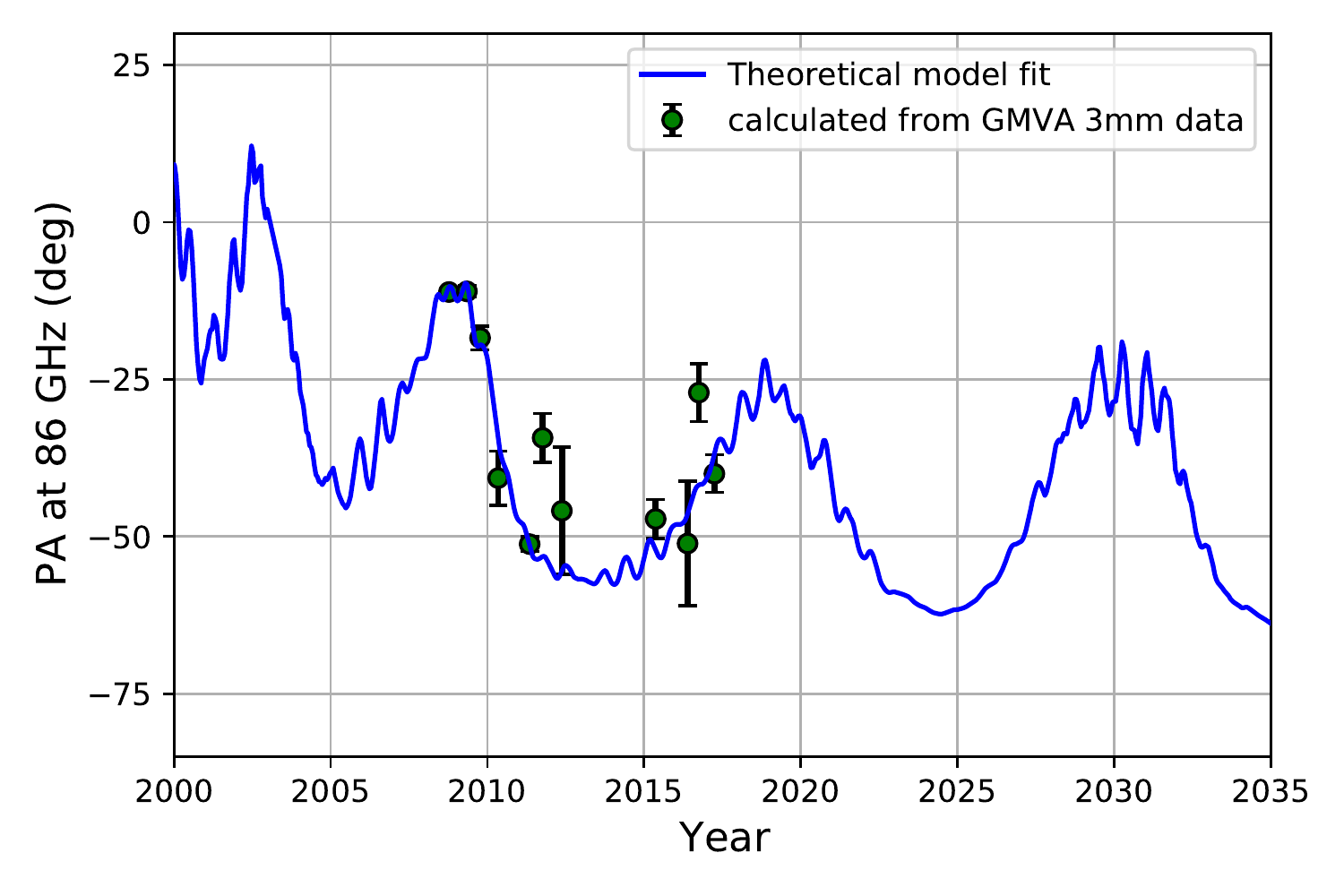}
    \caption{The position angle of the radio jet in 86 GHz observations (points) compared with the model line. The best-fit parameters are taken from Table 1 of \citet{Dey21} and we have used the disk model to calculate the theoretical model line. The viewing angle varies with time but is typically similar to the value found in \citet{hov09}. The delay time between the change of the jet direction at its origin and the appearance of the change in 86 GHz observations is 3.4 yr.}
    \label{fig:PA}
\end{figure}

OJ~287 has a prominent radio jet which is known for its large swings in the position angle in the sky \citep{Agu12}. These swings are well understood if they result from the variations in the orientation of the inner accretion disk, and if these variations are translated into the direction of the jet \citep{val13, Dey21}. Actually, these large swings were predicted by \citet{val11a} before they were observed. The high resolution data at 86 GHz is compared with a theory line in Figure~\ref{fig:PA} (see \citet{Dey21} for details). A 3.4 yr delay is found between the change of the disk axis and the direction of the jet. This is because the point at which the jet position angle is measured is typically 200 microarcsec from the radio core, and it takes some time to transit the information of the disk orientation change to this distance in the jet. The actual jet speed in OJ~287 is discussed e.g. in \citet{val12a}. In this case we find the relativistic $\Gamma ~\sim~ 7$. This is somewhat lower than the values measured by other methods. For example, \citet{sav10} give $\Gamma~\sim~9.3$, \citet{jor05} find $\Gamma~\sim~16.5$, and \citet{hov09} find $\Gamma~\sim~15.4$. Anyway, our $\Gamma ~\sim~ 7$ is close enough, and gives further support for the model. If the disk model was not accurate, or if the principles of jet-disk connection were not valid, there is no reason why the value coming out of this model would be anywhere near the values obtained via quite different methods.

\begin{figure}
    \centering
    \includegraphics[width=0.55\textwidth]{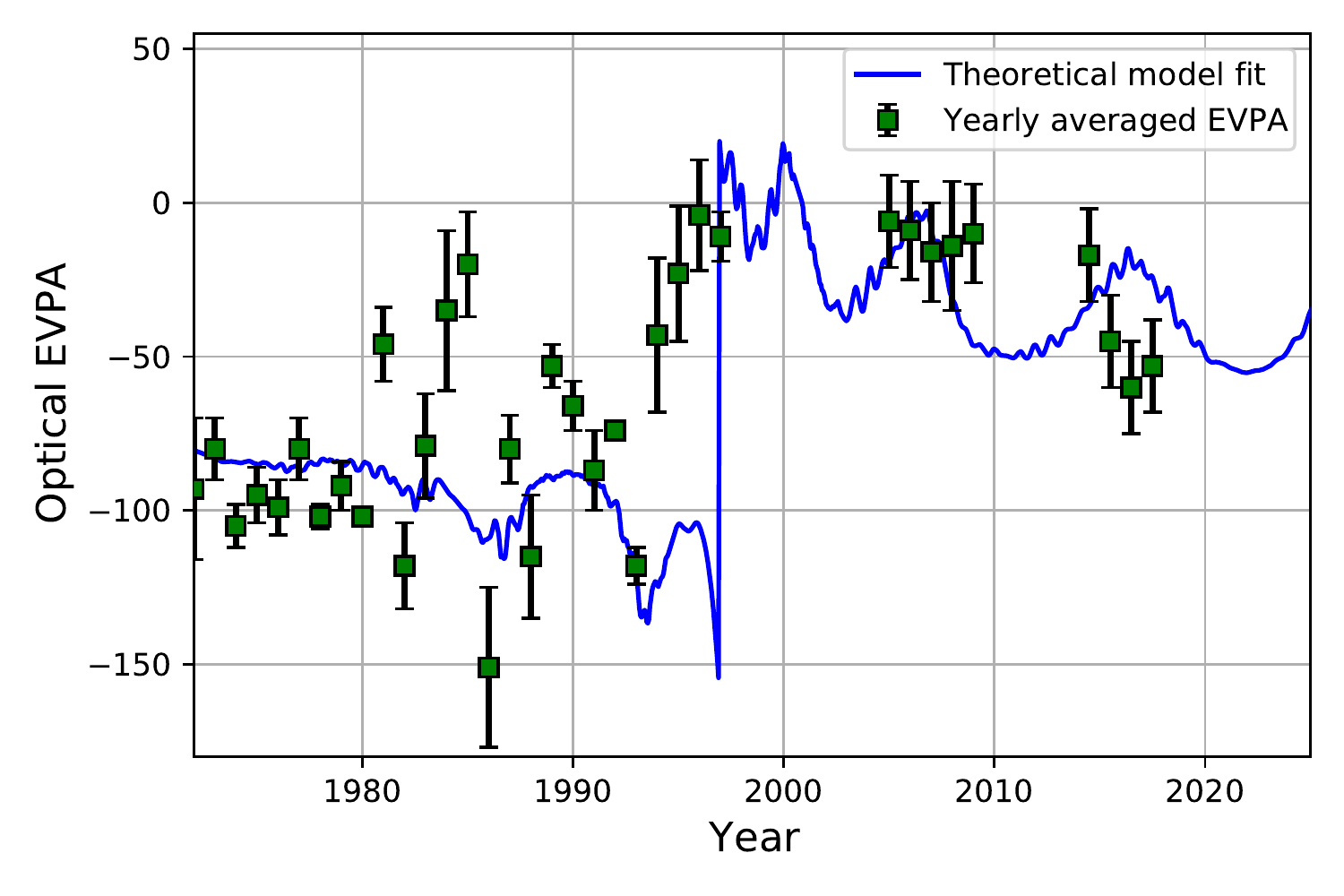}
    \caption{The annually averaged optical position angles of OJ~287 (points) together with the same jet model as in the radio jet observations. Note that the position angle has large fluctuations after the time of large optical flares, 1983/84 and 1994/95. The model parameters are the same as for the radio flares, except that time delay is 0.92 years, i.e. the optical emission region is closer to the center than the observed radio jet region at 86 GHz. This region cannot be resolved in optical observations.}
    \label{fig:EVPA}
\end{figure}

The region of the jet which we observe depends on the frequency because the resolution of the global VLBI maps is proportional to the observing frequency. Unfortunately this technique cannot be extended as far as the optical frequencies. However, some information about the jet may be obtained by monitoring of optical polarization. \citet{val12b} demonstrate that the optical polarization angle is a good proxy for the position angle of the radio jet. A big position angle jump occurred in 1995 in the optical polarization while a similar jump took place in 2009 in 15 GHz radio maps. Figure~\ref{fig:EVPA} shows how the optical polarization is modeled by our swinging jet model. The time delay is one quarter of the 86 GHz delay, so the distance of the region emitting in optical should be closer to the center by the same amount.

\begin{figure}
    \centering
    \includegraphics[width=0.6\textwidth]{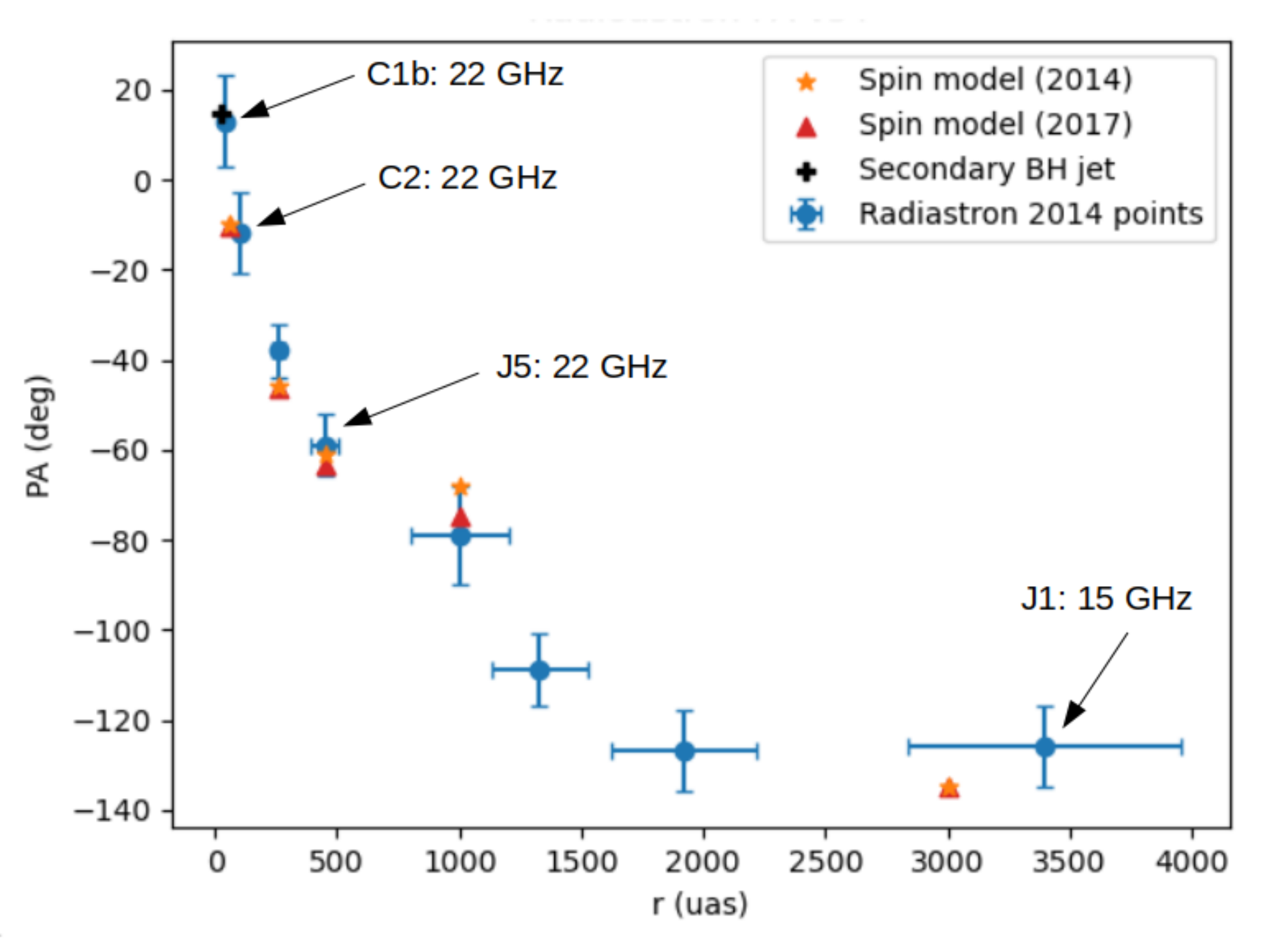}
    \caption{The distance and position angle PA of radio components in the Radioastron VLBI map by \citet{Gom21} with error bars. The 22 GHz data are used when available. The component J1 comes from 15 GHz observations.  Superposed are the theoretical positions of the corresponding components at two epochs (stars and triangles) in the spin model \citep{Dey21,val13}. The position of the expected secondary jet is also marked (cross). Its position angle is independent of epoch.}
    \label{fig:r_PA_radioastron}
\end{figure}

In the high resolution map from April/May 2014 \citep{Gom21}, the 86 GHz component C2 at 90 microarcsec from the center has the position angle $-8\pm45^{\circ}$. The latter component more or less coincides with the point in the jet where our model places the optical emission region (Figure~\ref{fig:r_PA_radioastron}). The innermost component C1b at 40 microarcsec lies at the position angle $13\pm10^{\circ}$, interestingly close to the theoretically calculated position angle the secondary jet at the position angle $14.6\pm4^{\circ}$ of \citet{Dey21} (see Figure~\ref{fig:r_PA_radioastron}).
Below we explore other possibilities of searching for observational evidence of the secondary black hole.


\section{On the presence of secondary BH accretion disk }

The mass ratio of the two black holes in the OJ~287 binary is 122. Thus for the same $\dot{m}$ and $\alpha$, the primary disk should be 122 times brighter. However, if the mass accretion rate is 122 times greater in the secondary than in the primary, both disks could be equally bright. This would put the brightness of the secondary over the Eddington limit, which may be possible at least in some accretion disk models \citep{hei07}.
It is well known that tidal perturbations increase the accretion flow considerably, see e.g. \citet{sil88}.

We may use \citet{sun97} and \citet{Pih13} to estimate the accretion rate to the secondary. In these simulations one particle represents $\sim 100$ solar mass of gas and the gas is accreted at the average rate of $\sim$ 5 particles per year, i.e. the mass accretion rate is about $5\times 10^2$ solar mass per year. This is approximately the mass contained within the Bondi-Hoyle accretion radius for the secondary, and much of it is likely to remain bound to the secondary when it leaves the disk. At what rate this matter is accreted to the black hole itself is another problem: it may well be that only some fraction of the gaseous envelope around the secondary black hole finds its way to the center before it suffers another collision with the disk. On the other hand, the steady state accretion rate of the primary is much lower, $\sim 0.6$ solar mass per year \citep{val19}. Thus the higher accretion rate to the secondary may compensate for its lower mass, in which case the secondary may not be all that much fainter than the primary.

However, the secondary is truncated by tides which may reduce its optical emission below the brightness of the primary, since the optical emission is associated primarily with larger radii in standard disk models. Also the accretion by the secondary is not steady, and its brightness may vary considerably depending on the epoch of observation. Not seeing it at one epoch does not mean that it could not be seen at all.

At present there are several ways by which the secondary could show up. In spectral observations there could be extra lines that match the redshift of the secondary at that particular instant in time. Because of the high orbital speed, these lines would be found quite far apart from the primary redshift in the spectrum. For example in November 2021 the secondary should have had redshift 0.38, in contrast to 0.307 in the primary. However, the relative redshift changes constantly, and moreover, the secondary spends a great deal of time behind the accretion disk and cannot be seen. After the December 3 crossing to the far side of the disk we should not see it for another ten years, and even after that only briefly. This is because precession now lines up the secondary orbit such that the pericentre is towards us, and the secondary spends only about $1/10$ of its orbital time in the pericenter part of the orbit.

\begin{figure}
    \centering
    \includegraphics[width=0.55\textwidth]{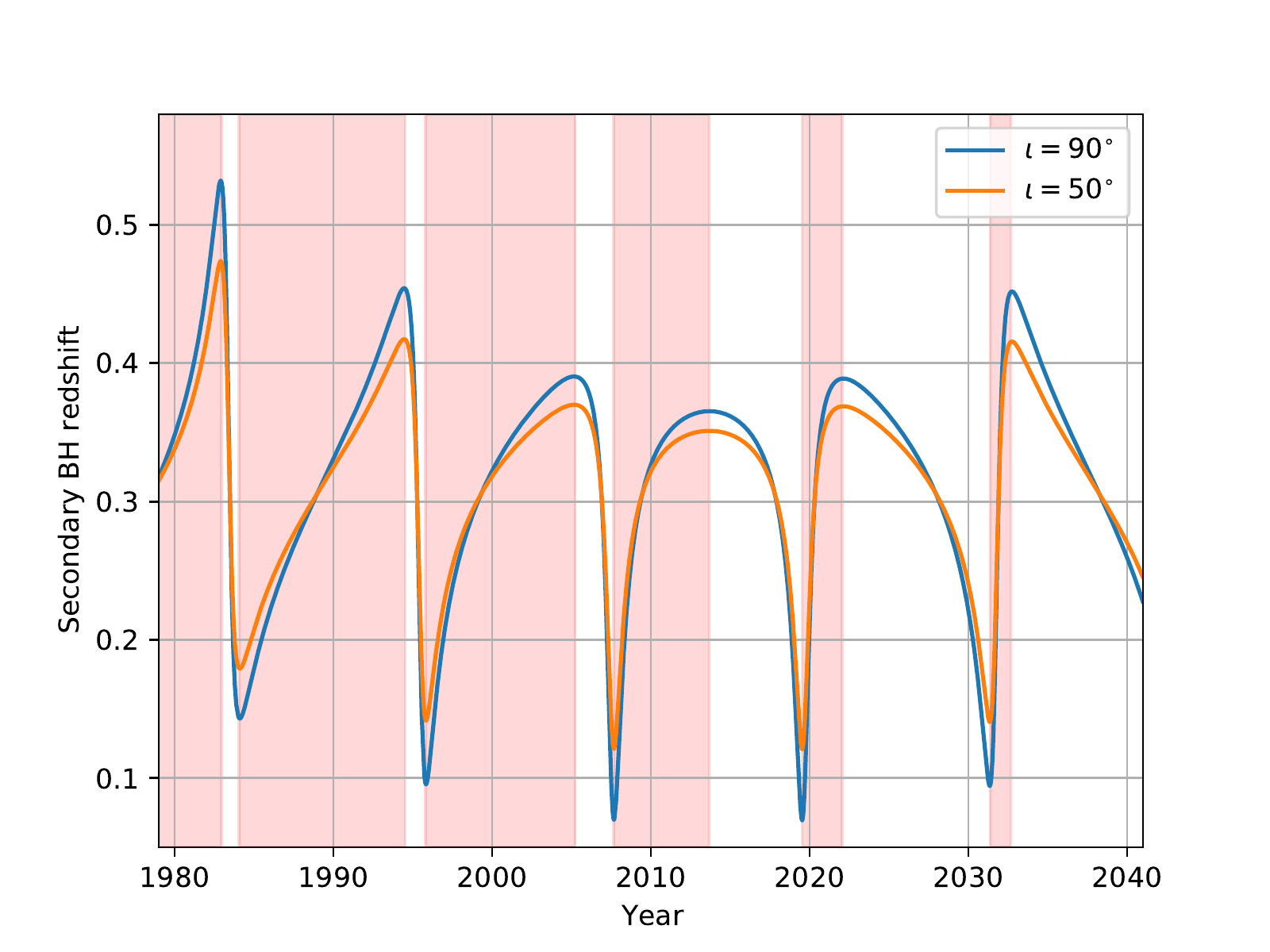}
    \caption{The redshift of the secondary black hole as a function of time. Two inclination values are shown: $90^{\circ}$ (edge-on) and $50^{\circ}$. The model does not specify the inclination. The shaded regions are the time intervals when the secondary is in front of the accretion disk and thus visible to us. The identification of the "front side" is based on taking the secondary to be in front of us during the 1989 deep fade \citep{tak90}}
    \label{fig:secondary_redshift}
\end{figure}

Table~\ref{tab2}, column 4, gives the expected wavelengths (in angstroms) of the secondary $H\beta$ or $H\alpha$ lines (the latter in square brackets). During the periods when the secondary should be behind the disk, we give the line for the opposite viewing direction, since we cannot be absolutely sure that we have identified the viewing direction correctly. Our interpretation in this respects relies heavily on \citet{tak90}. These values are in curved brackets, and refer to $H\alpha$. In Figure~\ref{fig:secondary_redshift}, we show the theoretically calculated redshift of the secondary BH.

\begin{figure}
    \centering
    \includegraphics[width=0.55\textwidth]{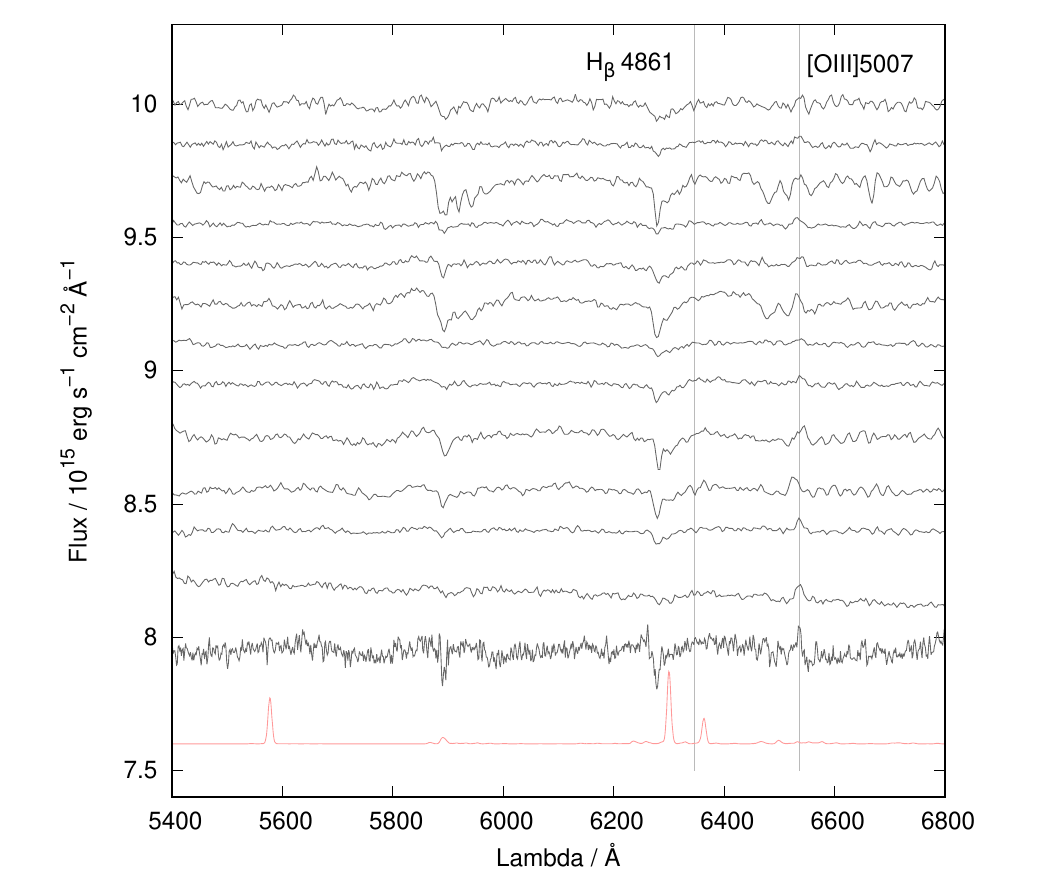}
    \caption{Optical spectra of OJ~287 taken at the {\it Nordic Optical Telescope}. The continuum has been fitted with a low-order polynomial and subtracted and the spectra have been shifted in vertical direction for clarity. The positions of $\lambda$4861 H$_{\beta}$ and $\lambda$50006 [OIII] at the redshift of OJ~287 are indicated by vertical lines. The red curve below shows the positions of prominent atmospheric emission lines. Apparent deep absorption structures in the spectrum are telluric lines. The epochs are (from top to bottom) : 12-02-2010, 23-10-2010, 18-11-2010, 04-01-2011, 17-01-2011, 01-10-2011, 11-02-2011, 23-04-2011, 07-05-2011, 16-11-2011, 19-02-2012, 27-12-2012, and 06-03-2013.}
    \label{fig:spectra}
\end{figure}

Among the VLT data, there is only one epoch where the expected secondary line would be within the observed wavelength range, and no line is seen. However, in this case we would need to have had the viewing direction wrong. Thus there is nothing in the VLT spectra that excludes the possibility of secondary lines in our current model. In the NOT and Rozhen spectra we are in the right wavelength region for $H\beta$, and the NOT spectra are taken at the time when we should see the secondary lines (see Figure~\ref{fig:spectra}). We do not see any of them. In fact, only a single primary line is detected weakly. Thus this non-detection does not give strong observational limits.  

The second way to find signs of the secondary is to look for the jet of the secondary. It may arise as soon as a new accretion disk has formed after a disk impact. Since the new accretion disk is formed from the material pulled out of the primary accretion disk, its axis should share the orientation of the axis of the mean disk of the primary. This is in contrast to the primary jet, which wobbles considerably around the axis of the mean disk under the influence of the companion \citep{Dey21}. A search for radio emission in a constant direction at the position angle $\sim15^{\circ}$ may reveal an indication of such a jet. The length of the jet should scale with the mass of the black hole, so that the secondary jet may be a factor $\sim 122$ shorter than the primary jet \citep{gho97,che15}. In the primary jet the outermost component J1 at 15 GHz is found at 3400 microarcsec from the center; thus the secondary jet may be expected in the scale of $\sim30$ microarcsec. This is the scale which has just been reached at VLBI \citep{Gom21} and will be studied in coming years.

Also, some of the flares may arise in the secondary jet rather than in the primary jet. They could be recognized for their short variability time scale. However, since the secondary jet is expected to represent only a fraction of the flux of the primary jet, the fast variations would appear at a low amplitude, consistent with \citet{val85} who found $1 - 10 \%$ peak-to-peak amplitude variations during the 1983/84 flaring season, with a 15.7 min period. 
\citet{Pih13} find that variations down to 3.5~d may be associated with the primary while the variability of the secondary could be faster by at least a factor of 122, depending on the secondary spin. For a maximally spinning secondary the 15.7 min period may arise \citep{Pih13a}. It is the period of the innermost stable orbit about the secondary black hole in the jet reference frame.

Since the secondary disk is likely to overflow its Roche lobe, we may also find interesting X-ray emission from the ring in the primary disk where the overflow stream terminates, akin to X-ray binaries. It is also worth searching for the direct accretion disk signature of the primary in the SED including a strong disk contribution in X-rays persistently detected in AGN with SMBHs of that mass.

\section{Are there more OJ~287-like systems?}

The remaining lifetime of the OJ~287 black hole binary before a merger is $10^4$ yr. Comparing this with the merger time scale of $10^7$ yr, there is only a chance of $\sim 10^{-3}$ to catch the system at this late stage of evolution. Is this a problem? It depends on the parent population of OJ~287-like systems. As an order of magnitude estimate, it has been deduced that there are something like $10^4$ binary black hole systems in the sky in the OJ~287 brightness category \citep{val10}. If this is the parent population, some ten of them would be found in the final stages of merger \citep{kro19}.

How could we find other OJ~287-like systems? We define an OJ~287-like binary black hole system such that it shows repeated bremsstrahlung flares due to the impacts on the accretion disk. The main requirement is that the mass ratio of the two black holes is rather large, so that the smaller black hole does not disrupt the accretion disk entirely in the process \citep{val12}. This is not very restrictive as large galaxies act as attractors of mergers with smaller galaxies, and together with the black hole mass - galaxy mass correlation this produces typically unequal pairs of black holes.

OJ~287 is a blazar with a jet pointing nearly toward us which increases its brightness many-fold, by about 3 magnitudes by our estimate above. The jets of other similar systems would likely point in some other direction. Thus it is unlikely that other OJ~287-like systems are found in lists of brightest radio sources, such as the Ohio Survey which gives the initial to the name of OJ~287. 

However, if such a source is discovered through the repeated bremsstrahlung flares, it will take time before enough flares are seen to solve the orbit. In the case of OJ~287 we are lucky that the source lies in the ecliptic zone which has been frequently photographed over 130 yrs. No such luck can be expected for the other OJ~287-like systems. It is also rather likely that these other systems contain smaller binary components by mass than OJ~287, with weaker bremsstrahlung flares. While other candidate binary black hole systems have been identified in recent years \citep[review by][]{KZ2016}, none of them was explained by involving optical bremsstrahlung flares, and none of them so far has the dense  polarimetry monitoring that OJ~287 has seen. Thus we may find more OJ~287-like systems, but not before full-sky monitoring programs have been operational for some time.
We now briefly list GW aspects of our description for OJ~287.

\section{Pulsar Timing Array response to nHz GWs from OJ~287's SMBH binary engine}
The above listed observational campaigns in many electromagnetic observational windows clearly point to the presence of a nHz GW emitting SMBH binary in OJ~287.
Therefore, it is important to develop prescriptions to model the expected pulsar timing array (PTA) response to nHz inspiral GWs from systems like OJ~287 \citep{Burke-Spolaor2019}. 
The fact that the orbital eccentricity and primary BH spin can substantially influence the BH binary orbital evolution, as demonstrated in \citep{Dey18,klein2018}, prompted us to employ an improved version of GW phasing approach to model temporally evolving GW polarization states from such systems \cite{damour2004,Susobhanan2020,dey2021b}.
This is required as the PTA relevant GW-induced pre-fit pulsar timing residual, namely  PTA response to inspiral GWs, induced by an isolated GW source at a given epoch $t$ may be written as 
\citep{Anholm2009} 
\begin{align}
    R(t) = \int_{0}^{t} \left[ h(t' - \tau_{P}) - h(t') \right]\  dt' \,,
    \label{eq:PTA_signal}
\end{align}
where $\tau_P$ is the geometric time delay in the  solar system barycenter (SSB) frame of the underlying millisecond pulsar while $t$ and $t'$ are the SSB relevant coordinate time variables.
Further, $h(t)$ provides the effective GW strain induced by an isolated source at an epoch $t$ and depends on the intrinsic parameters of the GW source like the total mass, spin values, orbital eccentricity and angles that specify its sky location, as detailed in  \cite{Susobhanan2020}.
In other words, PTA response of a pulsar to inspiral GWs from OJ~287 like system at a time $t$ depends not only on the strength of the inspiral GWs at the SSB at $t$ but also on the GW amplitude at location of the pulsar at time $t - \tau_P$. 
This is another reason why we need to employ PN approximation to track the orbital evolution and the resulting GW polarization states while modeling PTA response to inspiral GWs from SMBH binaries \citep{Susobhanan2020}.

In an ongoing effort, we are modeling temporally evolving quadrupolar order $h(t)$ from comparable mass BH binaries inspiraling along PN-accurate eccentric orbits under the influence of dominant order spin-orbit interactions due to the primary BH spin.
This is being done by adapting and extending the GW phasing efforts of \citet{damour2004,Susobhanan2020}.
It may be recalled that this approach usually employs Keplerian-type parametric solution to model the conservative dynamics of SMBH binaries in eccentric orbits.
Thereafter, the effects of GW emission are incorporated by adapting the method of variation of constants, as detailed in \citep{damour2004}.
In practice, we invoke the arguments of energy and angular momentum balance that lead to differential equations for orbital frequency and eccentricity with the help of orbital averaged PN-accurate far-zone energy and angular momentum fluxes \cite{dey2021b}.

\begin{figure}
    \centering
    \includegraphics[width=0.6\textwidth]{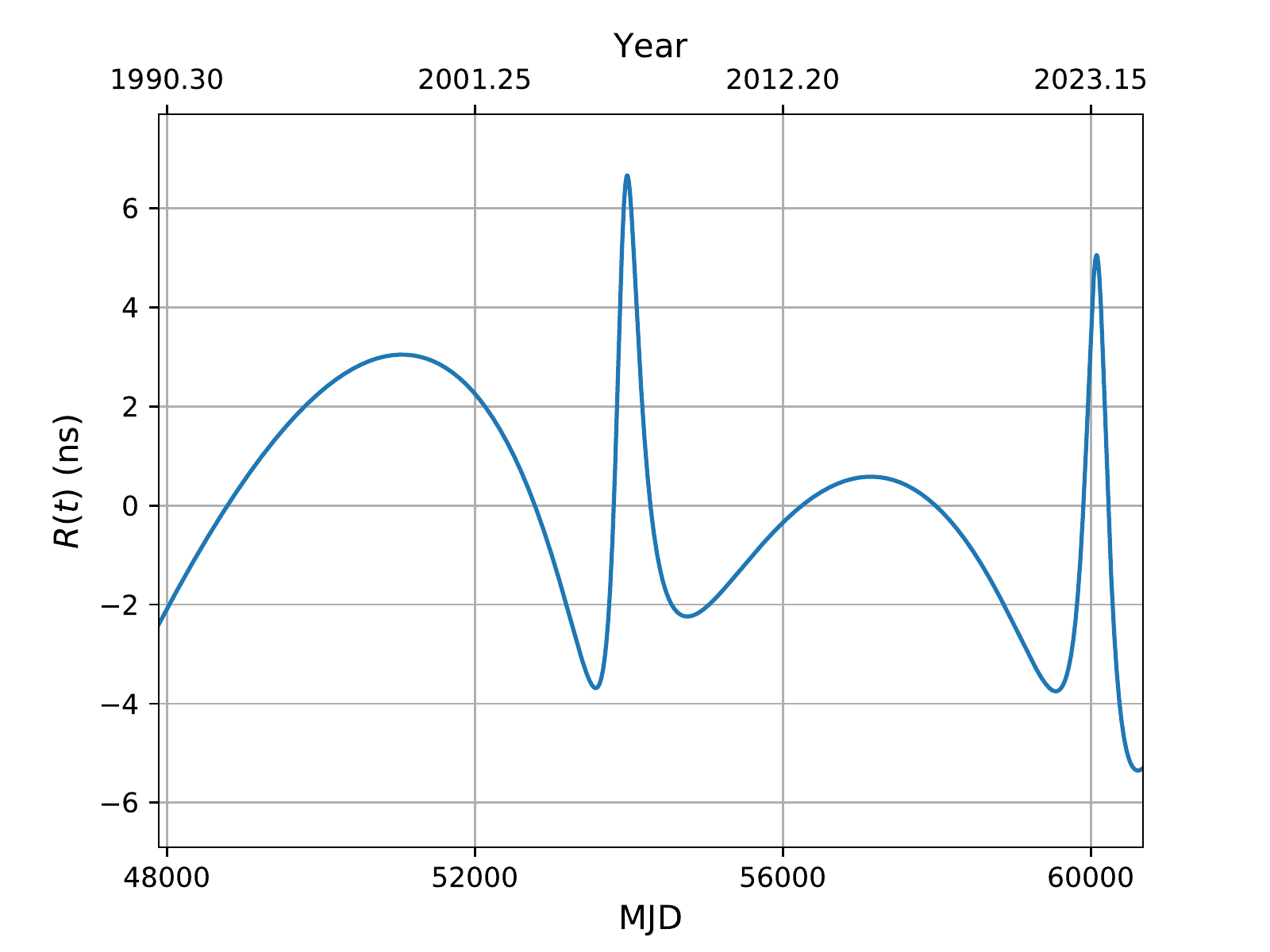}
    \caption{Timing residuals induced by inspiral GWs from OJ~287's SMBH binary on a typical PTA pulsar  PSR J0751+1807.
    We let the pulsar distance to be 1.4 Kpc \cite{IPTA_DR2} while SMBH binary parameters as listed in Section~\ref{subsec:BBH_model} with its luminosity distance to be 1.6 Gpc.
    Timing of milli-second pulsars with SKA are expected to achieve similar timing residuals. 
    }
    \label{fig:pta_signal}
\end{figure}

In Figure~\ref{fig:pta_signal}, we plot the expected PTA response of PSR J0751+1807 to inpsiral nHz GWs associated with our SMBH binary central engine description for OJ~287 as detailed in \citep{Dey18}.
The plot is based on Equation~\ref{eq:PTA_signal} where we have employed quadrupolar order $h(t)$ expression for spinning BH binaries in eccentric orbits from \citet{KG2005} while the orbital parameters are as detailed in \citet{Dey18}.
The sharp features depend on the orbital 
eccentricity and pulsar distance and provide unique signatures of the expected PTA response 
to nHz GWs from such SMBH binaries.
Unfortunately, the expected timing residuals are substantially below what we can achieve with the current timing campaigns on PSR J0751+1807.
However, it should be possible to find and time milli-second pulsars with such exquisite timing precision during the SKA era \cite{YiFeng20}.
Therefore, it should be possible to monitor persistent inspiral GWs from SMBH binaries like OJ~287 during the SKA era.
This development should naturally allow us to pursue persistent multi-messenger GW astronomy with crucial inputs from the ongoing, planned and proposed multi-wavelength electromagnetic observations of OJ~287.
This is because such observational campaigns  should allow us to constrain various astrophysical parameters of the binary BH central engine description for the unique blazar OJ~287.

\section{Discussion}

This article emphasizes the importance of continuing monitoring of OJ~287 at all wavelengths of the electromagnetic spectrum as well as in gravitational waves. Even though the binary orbit has now been solved \citep{Dey18} and confirmed \citep{Laine20}, there are many aspects of the system which still need verification and from which we can learn interesting astrophysics.

(1) The confirmation of the 2022 optical flare: This flare does not appear in any of the simple schemes that have been proposed over the years as alternatives to the LV96 binary black hole model. Difficult as it is to observe from the ground, we should just be able to observe the tail end of this flare. Polarisation observations play an important role here since the tail end of the flare should very highly polarised \citep{smi85,Villforth2010,val16}.

(2) The detection of the signals from the December 3, 2021 disk impact: This is energetically a huge event which has at previous times been studied only by occasional photometric measurements. This time we expect a full coverage from optical polarisation to X-rays which are the expected key indicators of the impact.
So far, in ongoing MOMO/Swift and ground-based monitoring observations, OJ 287 was found in a low-state in early December 2021 (Komossa et al., and Zola et al., in prep.). Also the timing of the impact depends on the detailed model which will be improved after the 2021/22 observing season.

(3) The $H\alpha$ flares: This has been caught only once, accidentally by \citet{sit85}. Later spectroscopic surveys, some of which are reported in this paper for the first time, have not been equally lucky. Now we give a rather precise prediction for the $H\alpha$ flare which follows the December 3, 2021, disk impact.

(4) The secondary black hole: The secondary black hole has never been directly detected. One of the problems is the short separation between the two black holes in the sky, typically 10 microarcseconds, depending on its orbital phase in an eccentric orbit. It will be several years before this resolution is achieved in radio VLBI. However, an attempt was made to detect the secondary by its spectral lines, but so far we only report a null result. It is a difficult task due to the high background level of the jet component. New attempts should be made during periods of low light level. Here we have made a prediction when such occasions may arise.

(5) Radio jet of the secondary: Even though the immediate surroundings of the secondary are hard to resolve, the current resolution in radio VLBI should be just enough to see the jet. A strong radio component was recently reported \citep{Gom21} at the expected position of the secondary jet \citep{Dey21}. It remains to be seen if this component behaves as expected. One problem is that the current direction of the primary jet is not very far from the suspected secondary jet. However, if the primary jet behaves according to our models \citep{Dey21}, the primary jet will move away from the secondary jet, and the identification becomes more clear. Thus the continuing radio VLBI monitoring of OJ~287 is essential.

(6) Radio jet of the primary: The radio jet of OJ~287 has been mapped in radio VLBI since 1980's. The origin of the jet and its connection to the central black hole and its accretion disk are still open problems. In the LV96 binary black hole model it is possible to calculate the direction of the spin axis of the primary as well as the normal to the innermost disk. They follow each other more or less \citep{Dey21}, with the disk normal showing a little more short time scale wobble than the spin axis. If we make the reasonable assumption that the initial jet direction is determined either by the spin axis (spin model) or the disk normal (disk model), we obtain a unique model for the jet which wobbles predictably. When projected onto the sky, we obtain position angles at any point in time. And since the resolution at higher frequencies is better than at lower frequencies in the global VLBI, we get snapshots of the jet at different distances from the center. After solving for the unknown viewing angles, we get a full picture of the bulk flow in the jet.

(7) Optical polarisation angle: It is usually assumed that the optical emission arises in the jet closer to the black hole than the radio emission. We have applied the same model for the optical polarisation that works for the radio position angle, and find that the region of optical emission is about four times closer to the central black hole than the inner 86 GHz emission knot. It is interesting also that the optical polarisation gives the direction of the innermost jet. This is not surprising since the jet has been found to have helical magnetic field geometry \citep{Gom21}. When compressed along the jet axis in shock fronts, the field lies practically perpendicular to the jet axis and the polarisation is lined up with the jet. It is important to continue regular polarisation monitoring programs to learn more about the inner jet.

(8) Optical fades: The times of low light level in the jet are useful also for the detection of the host galaxy. The absolute magnitude of the host has been measured by surface photometry out to $\sim 30$ kpc from the galaxy center, beyond which it is difficult to measure any light above the background generated by the bright point source, the AGN itself. Its brightness measured out to this distance is among the highest ever detected, and therefore it is likely that the total brightness of OJ~287 host is also among the highest. The brightness may also be measured by multicolor photometry at low light, because the color of the host is expected to be very different from the color of the jet. Thus the combined color of the source becomes more and more like galaxy colors when the source fades. This method suggests a brightness increase by one magnitude if the outer envelope of the host galaxy is included.

(9) High-energy emission and broad-band SEDs: Very dense, simultaneous X-ray--UV--optical monitoring has allowed us to identify the latest binary after-flares in 2017 and 2020, based on the brightest X-ray flares so far discovered from OJ 287 and other arguments \citep{Komossa2020}. Ongoing monitoring will reveal new outstanding X-ray--UV outbursts. Spectroscopy, SED modelling, cross-correlation analyses and structure function analyses \citep{Komossa2021b, Komossa2021d} will allow us to probe in detail the binary and outburst physics across the electromagnetic spectrum, and will allow to search for an accretion disk contribution of the secondary at select epochs. Ongoing high-energy monitoring could also serve as trigger criterion for new VHE observations.

(10) Very high energy gamma rays: VHE gamma rays have been detected in OJ287 only once. This occurred after the two big tidal flares which followed the 2015 impact flare. Even though we do not know the mechanism which produced this event, it may useful to monitor OJ~287 again after the tidal flares which follow the 2022 impact flare. More frequent monitoring would be also useful to see if the two phenomena are really connected. 

(11) Fast variability: The shortest time scale of periodic or quasi-periodic variability that may be associated with the primary jet is 3.5 days. This has been seen in many occasions. The time scale of variability should be $\sim 122$ times shorter in the secondary jet, due to the smaller mass of the secondary by this factor, or it could be even as low as $\sim 15.7$ minutes if the secondary spins much faster than the primary. Thus the study of fast variability in radio, optical and X-rays may reveal details of the secondary jet which are otherwise hard to recover.

12) Pulsar Timing Array: It will be important to develop accurate and efficient algorithms to model the expected PTA signals from SMBH binaries like OJ~287. Further, data analysis approaches that can efficiently look for such nHz GW sources in the PTA data sets of SKA era will need to be developed \cite{Burke-Spolaor2019}. Additionally, it will be desirable to search for milli-second pulsars close to the Sky location of OJ~287 as they are expected to be more sensitive to inspiral GWs from OJ~287. Investigations should be pursued to figure out the possibility of tracking the trajectory of secondary BH during the ngEHT era.

13) It has been argued that OJ~287 may produce a few muon neutrinos with energies more than 100 TeV in instruments like IceCube-Gen2 during ten years of Fermi flaring epochs, influenced by the modeling of the 2017 flare of TXS 0506+056 \citep{Oikonomou2019, Icecube2018}. The neutrino studies will be an important part of multi-messenger astronomy with OJ~287 in the coming decades.

\vspace{6pt} 



\authorcontributions{M.J.V. and L.D. have been responsible for the main design of this article as well as for theoretical developments. L.D and A.G. were in charge of the gravitational wave part, S.Z. of optical photometry, S.K. of the MOMO project and X-ray parts, T.P. and S.Z. of optical spectroscopy, R.H. of historical magnitude measurements, H.J. and A.V.B. of optical polarimetry and J.L.G. of radio mapping.}

\funding{
L.D. and A.G. acknowledge the support of the Department of Atomic Energy, Government of India, under project identification \# RTI 4002.
A.G. is grateful for the financial support and hospitality of the Pauli Center for Theoretical Studies and the University of Zurich.
S.Z. acknowledge the grant NCN 2018/29/B/ST9/01793.
R.H. acknowledges support by GACR grant 13-33324S and by the MSMT Mobility project 8J18AT036.
}

\acknowledgments{We thank Kari Nilsson for a valuable contribution in data reduction and producing one of the figures. We also appreciate the help that Vilppu Piirola has provided and continues to provide in optical polarisation observations of OJ~287. We also thank Gene Byrd and Shirin Haque for reading the manuscript and for valuable inputs.}

\conflictsofinterest{The authors declare no conflict of interest.} 






\reftitle{References}

\end{paracol}
\end{document}